S.E. KURATOV, I.S. MENSHOV, S.YU. IGASHOV, E.M. URVACHEV,

S.I. BLINNIKOV, D.S. SHIDLOVSKI, S.I. GLAZYRIN

Moscow, Dukhov Automatics Research Institute (VNIIA)

# NOVEL HYDRODYNAMIC CUMULATION MECHANISM CAUSED BY QUANTUM SHELL EFFECTS.

The computational and theoretical analysis carried out in this article demonstrates the existence of a nontrivial mechanism for the compression of a submicron-sized gas bubble formed by a gas of classical ions and a gas of degenerate electrons. This mechanism fundamentally differs from conventional compression mechanisms. It is shown that taking into account the quantum effect of a large spatial scale in the distribution of electrons qualitatively changes the character of cumulative processes. Because of a large-scale electric field caused by quantum shell effects, the compression process is characterized by the formation of multiple shock waves. The values of gas temperature and pressure achieved during compression occur higher by two orders of magnitude as compared with the classical adiabatic regime. The analysis is carried out within the framework of the following model: the dynamics of the electron subsystem is described by equations of a quantum electron fluid, while the hydrodynamic approximation is adopted for the ionic subsystem. The large scale effect is taken into account by means of effective external field acting on electrons. The theoretical analysis carried out within this approach clarifies the nature of the cumulative process in the system under consideration; some quantitative characteristics obtained with numerical simulation are presented. The possibility of experimental observation of this cumulative mechanism is analyzed. It is suggested that the manifestation of the effect can be observed during laser compression of a system of submicron targets by measuring the neutron yield

# Introduction

Previously, in Refs. [1] and [2], a new nontrivial manifestation of quantum effects in a spherical mesoscopic system of degenerate electrons (with a number of electrons $Ne \sim 10^4$-$10^9$) has been demonstrated. An unexpected absence of the simple tendency toward a uniform distribution of the electron density has been found, which has been tacitly assumed in the limit of large $Ne$. It has been shown that the electron density has a large spatial scale which is of the order of the system size and is much larger than another spatial scale - the Fermi wavelength of the electron gas. This large scale effect clearly manifests itself in the appearance of an electric field acting on the ionic system. The spatial distribution of the potential has an oscillating behavior with several extrema. This result has been obtained using four different approaches: (a) the analytical method of semiclassical Green's functions, (b) the direct numerical summation over the occupied electron states in the spherical cavity with impenetrable wall, (c) the numerical calculations based on the method for constructing the Green's function by means of two linearly independent solutions to the Schrodinger equation and (d) the density functional method (DFT). The results obtained using these approaches are found to be in good qualitative and quantitative agreement. This effect is demonstrated [1, 2] by using the following two systems: (i) free degenerate electrons in a spherical well and (ii) electrons in a compressed gas bubble. The spherical symmetry of the system or proximity to it is of fundamental importance for the appearance of the effect.

One of the possible manifestations of this effect is a non-trivial mode of compression of gas bubbles containing a mixture of a gas of degenerate electrons and a gas of classical ions, which is fundamentally different from the conventional gas compression. This circumstance has been first pointed out in [1]. In the present work, we give a detailed analysis of the hydrodynamic processes occurring during the compression of a submicron-sized gas bubble taking into account the effect of a large-scale spatially oscillating electric field.

The significant computational complexity is associated with the computational-theoretical analysis in the direct formulation, which requires self-consistent calculation the gas-dynamic motion of ions and solving the quantum mechanical problem for determining the spatial distribution of the mesoscopic number of electrons immersed into the medium of ions. Thus it turns out to be necessary to elaborate a simplified model formulation. Its essence is as follows. The ionic subsystem is described by the equations of hydrodynamics in the presence of an external electrostatic force. This electrostatic force is determined from the equations of quantum hydrodynamics for the electronic subsystem, in which an effective field is introduced to take into account the large-scale effect. A simple analytical expression with variable coefficients is used as an approximant of numerical results for the effective field.

The model proposed is employed for the computational-theoretical analysis of the problem on the compression of a submicron gas bubble. This study shows that because of the large-scale oscillating electric field caused by quantum shell effects, the compression process is characterized by the formation of multiple shock waves in the gas bubble, which qualitatively changes the nature of cumulative processes. The values of gas temperature and pressure achieved in the process of compression increase by two orders of magnitude as compared to the conventional regime of compression. This result is also confirmed in numerical simulations.

Detecting this effect can be attempted measuring neutron yield in experimental studies of laser compression of submicron targets. The present study analyzes one of possible experimental settings similar to modern laser-driven inertial confinement fusion (ICF) experiments carried out at the NIF facility [20]. For example, a system of $10^4$ nanometer targets is placed in a Hohlraum and each of the targets is compressed by X-rays. When each target absorbs energy on the order of $10^{-13}$ J, it is estimated that 100 neutrons are generated in the system. If there is no effect in the system, neutrons should not be generated. An energy of the order of $10^{-11}$ J is spent on the production of one neutron. This is comparable with that obtained in recent experiments at NIF [3–4], in which a record neutron yield of $10^{17}$ has been measured at an input energy of $10^6$ J.

# 1. Formulation of the model

## 1.1 Mathematical model

The object to be studied is a submicron-sized bubble formed by a gas of degenerate electrons ($\varepsilon_F > 0.1 kT$) and a gas of classical ions. Such conditions can be realized in the range of pressure values $p \sim 0.1\text{-}2$ TPa and temperature $T < 1$ eV.

Under these assumptions, the radius of the bubble $R$ satisfies the following inequality:

$$\frac{R}{V_F^e} \ll \frac{1}{N^{\frac{1}{3}} V_i}$$

where $V_F^e$ is Fermi velocity of electrons, N is number of electrons, $V_i$ is ion velocity. This condition provides instant adjustment of the electronic subsystem to the profile of the ionic subsystem. That is, each current profile of the spatial distribution of ions corresponds to a certain spatial distribution of electrons. The emerging nonzero electric field acts on the ionic subsystem as a spatially distributed force. The ionic subsystem is compressing under this force and also an external force applied to the outer surface of the bubble. Since the mean free path of ions is much less than the characteristic size of the system, their motion can be described in the hydrodynamic approximation taking into account an external force that depends on the ion concentration.

The problem is studied at a high concentration of particles $n_e = 10^{24}$ cm$^{-3}$ or more so that the hydrogen atoms (to be considered) are so close to each other that the screening radius becomes smaller than the Bohr radius and the bound electron states disappear [5]. Therefore, under such conditions, hydrogen is completely ionized and is in the metallic state (Mott transition). This justifies the calculation of electron density distribution in the framework of the DFT method using the "jelly" model in which ions are approximately represented as a continuous homogeneous distribution of positive charge and electrons are not bound to ions as it is in ordinary metals. Thus, to simulate accurately the compression of the gas bubble, it is necessary to solve self-consistently the following three problems (self-consistent formulation of the model):

- the quantum mechanical problem of determining the distribution of electrons against the background of ions by the DFT method;

- determining the emerging electric field which forms a force acting on the ionic subsystem by solving the Laplace equation;

- the hydrodynamic problem of the motion of the ionic subsystem in the presence of the force.

Such statement of the problem is associated with significant computational complexities and requires vast computational resources due to the quantum mechanical part deals with solving the Schrödinger equations for the mesoscopic number of electrons $N_e \sim 10^9$. The computation time in numerical simulations drastically increases as soon as it becomes necessary to make repeated calculations of the quantum mechanical task $\sim 10^3$ to $\sim 10^4$ times per one gas-dynamic step to ensure the required accuracy. Therefore, we use a simpler problem statement where the hydrodynamic equations of a quantum electron fluid [6] are solved for the evolution of the electron subsystem rather than the quantum mechanical problem. Such an approach, for example, is successfully used in modeling a quantum electron gas in plasma [21–25] and in various semiconductor electronic devices [7–9]. The validity of this approach for the system analyzed in the present work is discussed below.

The mathematical model to be considered consists thus of the following three parts.

1. The system of equations that determines the distribution of electrons looks as follows:

$$\frac{\partial \rho_e}{\partial t} + \frac{1}{r^2}\frac{\partial}{\partial r}\left(r^2 \rho_e v_e\right) = 0 \quad , \qquad (1)$$

$$\frac{\partial v_e}{\partial t} + v_e \frac{\partial v_e}{\partial r} = -\frac{1}{\rho_e}\frac{\partial p_e}{\partial r} - \frac{e}{m_e}\frac{\partial}{\partial r}(\varphi + U_{eff}) \quad , (2)$$

$$\frac{de_e}{dt} + p_e \frac{d(1/\rho_e)}{dt} = 0 \quad . \qquad (3)$$

Here $r$ is the distance from the origin, located in the center of the spherical bubble, $\varphi$ is the electrostatic potential, $p_e$ is the quantum-statistical pressure of the degenerate electron gas and $e_e$ is its internal energy, $\rho_e$ and $v_e$ are density and velocity, $e$ and $m_e$ are charge and mass of electron. Fundamentally, new feature of this model is the presence of an effective field $U_{eff}$ in the equations of motion for the electrons, which models the large-scale effect. It is worthwhile to split the effective potential in two terms: $U_{osc}(r)$ and $U_{bar}(r)$.

The terms introduced have the following meaning: the first one $U_{osc}$ simulates the large scale effect in the bulk of the volume of the bubble while the second $U_{bar}$ is the barrier potential associated with the description of the electron subsystem by the equations of a quantum liquid which assumes that the characteristic spatial scales of changes in the electron density are much larger than the inter-electron distance. The electron density sharply drops down to zero near the bubble boundary at a distance of the order of several De Broglie wavelengths of electrons in accordance with the zero boundary condition imposed on the wave functions of electrons. Thus, the quantum behavior of the wave functions of electrons in the quantum fluid formalism we model by introducing a barrier potential. A detailed analysis of the possibility of this approach is given below.

It should be noted that the Bohmian term

$$\varphi_{Bohm} = \frac{h^2}{2m_e\sqrt{n_e}}\nabla^2\sqrt{n_e}$$

in the above equation of motion (2) for the electron quantum liquid was neglected due to its smallness compared to $U_{eff}$.

2. The Poisson equation which determines the force acting on the ionic system:

$$\Delta\varphi = \frac{e}{\varepsilon_0}(n_i - n_e) \quad . \qquad (4)$$

3. The system of equations describing the ionic subsystem has the form:

$$\frac{\partial \rho_i}{\partial t} + \frac{1}{r^2}\frac{\partial}{\partial r}\left(r^2 \rho_i v_i\right) = 0 \quad , \qquad (5)$$

$$\frac{\partial v_i}{\partial t} + v_i \frac{\partial v_i}{\partial r} = -\frac{1}{\rho_i}\frac{\partial p_i}{\partial r} - \frac{e}{m_i}\frac{\partial}{\partial r}\varphi \quad , \quad (6)$$

$$\frac{de_i}{dt} + p_i \frac{d(1/\rho_i)}{dt} = 0 \quad . \quad (7)$$

Where the notation is analogous to the system (1)-(3) and the subscript $i$ indicates parameters related to ions.

The system of equations (1)-(7) is greatly simplified as applied to the problem under study. We consider the compression of the gas bubble in the neutral condition when $n_i \cong n_e$ at any moment as the characteristic time of the motion of the ionic subsystem is much greater than the relaxation time of the electronic subsystem. Within this approximation, the system of equations (1)-(7) can be rewritten as follows

$$0 = -\frac{1}{\rho_e}\frac{\partial p_e}{\partial r} - \frac{e}{m_e}\frac{\partial}{\partial r}(\varphi + U_{eff}) \quad (8)$$

$$\frac{\partial \rho_i}{\partial t} + \frac{1}{r^2}\frac{\partial}{\partial r}\left(r^2 \rho_i v_i\right) = 0 \quad (9)$$

$$\frac{\partial v_i}{\partial t} + v_i \frac{\partial v_i}{\partial r} = -\frac{1}{\rho_i}\frac{\partial p_e}{\partial r} - \frac{1}{\rho_i}\frac{\partial p_i}{\partial r} - \frac{e}{m_i}\frac{\partial}{\partial r}(U_{eff}) \quad (10)$$

$$\frac{de_i}{dt} + p_i \frac{d(1/\rho_i)}{dt} = 0 \quad (11)$$

In subsequent sections the system of equations (8)-(11) is used for the analysis of cumulative processes during the compression of a submicron size gas bubble.

1.2 Modeling effective potential $U_{eff}$:

Quantum shell effects in the system under consideration appear as large-scale inhomogeneity in the spatial distribution of the electron gas. The proposed formalism of quantum hydrodynamics is intended for taking into account of such features of spatial distribution in description of a mesoscopic system of degenerate electrons. This is achieved by means of introduction of an external effective potential $U_{eff}$ in the equations of electron motion (8):

$$\frac{\partial v_e}{\partial t} + v_e \frac{\partial v_e}{\partial r} = -\frac{1}{\rho_e}\frac{\partial p_e}{\partial r} - \frac{e}{m_e}\frac{\partial}{\partial r}(\varphi + U_{eff}) \quad .$$

Accordingly, for the model considered, it is vitally important to determine correctly the shape of this potential.

The dependence of the potential on the radius $r$ can be determined in terms of the density $n_e^{free}$ of free degenerate electrons in the potential well that is calculated within the DFT method. The averaged distribution of the electron density $\langle n_e^{free} \rangle$ is then calculated based on $n_e^{free}$ by means of smoothing procedure that smears out small scale density fluctuations (on scales of the order of the Fermi length) while preserving large scale structures. The distribution $\langle n_e^{free} \rangle$ found in this way allows to calculate $U_{eff}$ with the aid of the following relation:

$$\frac{1}{\langle n_e^{free} \rangle} \frac{\partial p_e(\langle n_e^{free} \rangle)}{\partial r} = - e \frac{\partial}{\partial r}(U_{eff}(r)) \quad , \quad (12)$$

which results from Eq. (8) in the absence of an electric field, i.e., for φ=0.

The smoothing procedure can be carried out in various ways detailed as follows. The first one was used successfully in [1]–[2] is to calculate the electric potential of an electrically neutral system as a whole consisting of a non-interacting gas of free degenerate electrons and an ionic spatially homogeneous ball. The potential is conventionally calculated by numerical integration of the charge density represented by a set of values in the grid nodes with the Green's function of the Laplace equation. After that the averaging is performed. Then, using the averaged potential, we determine the electron number density from the Poisson equation. As a result, the electron concentration averaged over small-scale fluctuations is obtained.

The second one, a more straightforward method, consists in direct averaging of the density found in the framework of a quantum mechanical calculation. Appendix 1 demonstrates the equivalence of the two averaging approaches and presents results of calculating $U_{eff}(r)$ within the second method. Figures A1.7-A1.10 represent the calculated values of the effective potential depending on the dimensionless factor $k_f R_0$ and normalized by the parameter $\frac{(3\pi)^{2/3}}{2^{5/3}} \frac{h^2}{mR_0^2}$, where $k_f$ is $\frac{p_f}{h}$, $h$ is Planck constant, $m$ is mass of electron.

The obtained results allow us to draw the following conclusion about the nature of the behavior of $U_{eff}(r)$. The potential consists of two qualitatively different terms,

$$U_{eff}(r) = U_{osc}(r) + U_{bar}(r) \qquad (13)$$

where $U_{osc}(r)$ is the effective oscillating potential acting in the almost bulk of the inner region of the bubble ($r<R_0$) excluding thin "skin" – a vicinity of its border $r=R_0$, while $U_{bar}(r)$ is the effective barrier potential. On the contrary, the latter is constant with high accuracy in the almost bulk of the inner region of the bubble ($r<R_0$) and is sharply changing in a narrow boundary region ($r\sim R_0$).

The shape of the potential $U_{osc}(r)$ has an oscillating character; it depends on the number of particles and has from 1 to 3 bumps. A certain periodicity in the shape of the potential can be observed, considering its dependence on the number of particles with a characteristic period approximately equal to $\Delta k_f R_0 \sim$ (Fig. A1.7-Fig/ A1.10) The amplitude of the effective potential is proportional to $\frac{1}{R_0^2}$ and weakly depends on $N$.

The shape of the potential $U_{bar}(r)$ has a stepwise profile and practically does not depend on the number of particles. The amplitude of the potential $U_{bar}(r)$ is proportional to $\frac{1}{R_0^2}$.

Appendix 1 presents the calculated electrostatic potential $\varphi_{electr}^{free}$ that arises in the electrically neutral system of a gas of degenerate non-interacting electrons and a spatially homogeneous ball of ions. Figures A1.13-A1.15 show the results of averaging $\varphi_{electr}^{free}$. The averaged potential qualitatively coincides with $U_{eff}(r)$ ; it depends on the number of particles and has from 1 to 3 local bumps.

In Appendix 2 fitting of the numerical effective potential by a simple analytical expression is developed. It has the following form:

$$U_{bar}(r) = \frac{W}{1+exp\left(\frac{r-R_{ws}}{a}\right)}, \quad U_{osc}(r) = \sum_{m}^{M} C_m \cos\left(\pi m \frac{r}{R_0}\right) ,$$

where $R_{ws}$, $C_m$ and $a$ are constants. These approximations are used in numerical simulations in Section 2.

Using these results, we show that these approximations make it possible to obtain satisfactory quantitative results for the amplitudes of the main functions obtained in [1] with extended calculations by the DFT method.

We introduce the following quantities characterizing the system considered:
$U_{osc}(r)$-- effective oscillating potential;
$\varphi_{electr}^{free}$ - electric potential of electrically neutral system composed of non-interacting electrons and a spatially homogeneous ball of ions;
$\varphi_{electr}^{interect}$ - the same as above except that electrons are considered to be interacting;
$\delta n^{free}(r)$ - deviation from the average value of the electron density ;
$\delta n^{interect}(r)$ - the same as above except that electrons are considered to be interacting.

Then, we proceed as follows. Using the simplest model for the electrostatic potential $\varphi_{electr}^{free} = \varphi_{0\ electr}^{free} \sin\left(\alpha \frac{r}{R_0}\right)$, $\alpha = 3\pi$ and the results of [1], we determine an analytical approximation for the amplitude $U_0^{osc}$ of $U_{osc}(r)$ in the following way.

Assuming that in the bulk of the volume, the deviation of the electron density from the average value is small, we obtain:

$$\langle n_e^{free} \rangle = n_0 + \delta n(r),$$

$$p_e = \frac{(3\pi^2)^{\frac{2}{3}}}{5} \frac{\hbar^2}{m_e} n^{\frac{5}{3}} = \frac{(3\pi^2)^{\frac{2}{3}}}{5} \frac{\hbar^2}{m_e} n_0^{\frac{5}{3}}\left(1 + \frac{5}{3}\frac{\delta n}{n_0}\right).$$

Substituting the resulting expression into Eq. (12), we find the relation connecting the functions $\delta n(r)$ and $U_{osc}$:

$$-eU_{osc}(r) = \frac{(3\pi^2)^{\frac{2}{3}}}{5}\frac{\hbar^2}{m_e}n_0^{\frac{2}{3}}\frac{5}{3}\frac{\delta n(r)}{n_0} \quad . \tag{14}$$

Using the Poisson equation

$$\frac{1}{r^2}\frac{\partial}{\partial r}\left(r^2 \frac{\partial \varphi_{electr}}{\partial r}\right) = \frac{e}{\varepsilon_0}\delta n,$$

one can get

$$-\varphi_{0\ electr}^{free} \sin\frac{\alpha}{R_0}\left(\frac{2r\sin\left(\alpha\frac{r}{R_0}\right)+r^2\frac{\alpha}{R_0}\cos\left(\alpha\frac{r}{R_0}\right)}{r^2}\right) = \frac{e}{\varepsilon_0}\delta n$$

From Eq. (14), we can then obtain the following expression for the amplitude of the oscillating part:

$$U_0^{osc} = \varphi_{0\ electr}^{free} \frac{(3\pi^2)^{\frac{2}{3}}}{3}\frac{h^2}{e^2 m_e}\frac{\varepsilon_0}{n_0^{\frac{1}{3}}}\left(\frac{\alpha}{R_0}\right)^2.$$

Taking into account that according to DFT calculations $\varphi_{0\ electr}^{free} = 0.02\frac{eN^{0.45}}{4\pi\varepsilon_0 R_0}$ [1], we finally get the following expression for the effective potential amplitude:

$$U_0^{osc} = 0.06\frac{h^2}{e\,m_e}\frac{N^{0.117}}{R_0^2}\frac{(3\pi^2)^{\frac{2}{3}}}{12\pi}\left(\frac{4}{3}\pi\right)^{1/3}\alpha^2 \quad (15)$$

This value is in good qualitative and quantitative agreement with the results obtained in Appendix 1 and Appendix 2.

Similarly, within the framework of the proposed approach Eqs. (8)-(11) with the aid of Eq. (15), one can calculate the amplitudes of other characteristic functions $\varphi_{electr}^{interect}$, $\delta n^{free}(r)$, and $\delta n^{interect}(r)$:

$$\delta n^{interect}(r) = 0.06\frac{N^{0.117}\alpha^2}{4\pi R_0^2}n_0^{\frac{1}{3}},$$

$$\varphi_{0\ electr}^{interect} \sim 0.02\frac{eN^{0.45}\alpha^2}{4\pi\varepsilon_0 R_0\left(\frac{4}{3}\pi\right)^{1/3}},$$

$$\delta n^{free}(r) \sim 0.06\frac{N^{0.45}\alpha^2}{4\pi R_0^3}.$$

These expressions also are in good agreement with the values obtained earlier in Refs. [1]-[2] for the amplitudes of the corresponding functions (see Figs. 4, 9 in [1]).

Thus, it is shown that the approach based on the hydrodynamic description of an electron gas with taking into account the effective potential allows to describe the main characteristic features of the system obtained within the DFT method. This proves the validity of the proposed approximate approach for the analysis of the system considered in the present article.

# 2. Peculiarities of hydrodynamic cumulation caused by quantum shell effects.

In this section, a new mechanism of hydrodynamic cumulation initiated by quantum shell effects is theoretically investigated within the framework of the model formulated above. We consider two different problems that demonstrate the basic mechanism of shock wave formation.

The first problem is related to "relaxation process" in a gas bubble being initially at rest when the effective potential is instantaneously switched on.

The second problem is "adiabatic compression process" of a gas bubble where the densities of electrons and ions are assumed to practically coincide and have initially a nonmonotonic spatial density profile with several local extrema. Any of the possible mechanisms of gas bubble compression can be reduced to a combination of these two processes.
By theoretical analysis, we demonstrate the causes of forming multiple shock waves and discuss conditions for the occurrence of this phenomenon. The numerical analysis demonstrates quantitatively the large-scale effect on the temperature values achieved in the cumulation.

2.1 The process of relaxation
The first problem is related to the relaxation process in the system with non-moving boundary being initially at rest with ($n_e = n_i = n_0$=const), in which the effective potential is instantaneously switched on. Such a situation can be realized, for example, under impact action on a metal cluster, in which the initial crystal lattice of ions is transformed into a plasma state. The final result of the process is formation of the static equilibrium ($v_i=0$) state of the system.
Relaxation process in the system is governed by the set of the following equations:

$$0 = -\frac{1}{\rho_e}\frac{\partial p_e}{\partial r} - \frac{e}{m_e}\frac{\partial}{\partial r}(\varphi + U_{eff}),$$

$$\frac{\partial \rho_i}{\partial t} + \frac{1}{r^2}\frac{\partial}{\partial r}\left(r^2 \rho_i v_i\right) = 0,$$

$$\frac{\partial v_i}{\partial t} + v_i \frac{\partial v_i}{\partial r} = -\frac{1}{\rho_i}\frac{\partial p_i}{\partial r} - \frac{1}{\rho_i}\frac{\partial p_e}{\partial r} - \frac{e}{m_i}\frac{\partial}{\partial r}(U_{eff}),$$

$$\frac{de_i}{dt} + p_i \frac{d(1/\rho_i)}{dt} = 0,$$

$$p_e = \frac{(3\pi^2)^{\frac{2}{3}}}{5}\frac{h^2}{m_e} n^{\frac{5}{3}},$$

$$p_i = n_i k T_i.$$

The first order perturbation theory approximation reads:

$$n(t) = n_0 + \Delta n(r,t),$$

$$v_i = \Delta v_i(r,t),$$

$$T_i = T_i^0 + \Delta T_i(r, t),$$

$$p_e = \frac{(3\pi^2)^{\frac{2}{3}}}{5}\frac{h^2}{m_e}n^{\frac{5}{3}} = \frac{(3\pi^2)^{\frac{2}{3}}}{5}\frac{h^2}{m_e}n_0^{\frac{5}{3}}(1 + \frac{5}{3}\frac{\Delta n}{n_0}),$$

as a result, the equations that govern the evolution of the system take the form

$$\frac{\partial \Delta n(r,t)}{\partial t} + \frac{\partial}{\partial r}\left(n_0 \Delta v_i(r,t)\right) + \frac{2}{r}\left(n_0 \Delta v_i(r,t)\right) = 0, \qquad (16)$$

$$\frac{\partial \Delta v_i}{\partial t} = -\frac{k}{m_i}\frac{\partial \Delta T_i(r,t)}{\partial r} - \frac{kT_i^0}{m_i n_0}\frac{\partial \Delta n(r,t)}{\partial r} - \frac{1}{m_i n_0}\frac{(3\pi^2)^{\frac{2}{3}}}{3}\frac{h^2}{m_e}n_0^{\frac{2}{3}}\frac{\partial \Delta n(r,t)}{\partial r} - \frac{e}{m_i}\frac{\partial}{\partial r}(U_{eff}),$$

(17)

$$\frac{3k}{2m_i}\left(\frac{\partial \Delta T_i}{\partial t}\right) - \frac{kT_i^0}{m_i n_0}\left(\frac{\partial \Delta n}{\partial t}\right) = 0 . \qquad (18)$$

The initial conditions are $\Delta n(r, 0) = 0$, $\Delta v_i(r, 0) = 0$, $\Delta T_i(r, 0) = 0$.

The inhomogeneous wave equation for $\Delta n(r, t)$ follows from Eqs. (16)-(18) straightforwardly:

$$\frac{\partial^2 \Delta n}{(\partial t)^2} - \frac{c_0^2}{r}\frac{\partial^2 (r\Delta n)}{(\partial r)^2} = \frac{en_0}{m_i}\frac{1}{r}\frac{\partial^2(rU_{eff})}{(\partial r)^2} . \qquad (19)$$

where

$$c_0^2 = \frac{5kT_i^0}{3m_i} + \frac{1}{m_i}\frac{(3\pi^2)^{\frac{2}{3}}}{3}\frac{h^2}{m_e}n_0^{\frac{2}{3}} .$$

Let us first consider the relaxation dynamics of the system caused by the oscillation potential, for which we adopt the simplest model form:

$$U_{osc}(r) = \frac{\beta}{(R(t))^2}\cos(\alpha\frac{r}{R(t)})$$

$$\beta = 0.02\frac{h^2}{e\, m_e}\sqrt[4]{N}\frac{(3\pi^2)^{\frac{2}{3}}}{12\pi}\left(\frac{4}{3}\pi\right)^{1/3}\alpha^2$$

Taking into account the initial conditions, the solution to Eq. (19) has the form

$$\Delta n(r, t) = \frac{n_0 e}{m_i c_0^2}\frac{\beta}{(R_0)^2}\cos\left(\alpha\frac{r}{R_0}\right)\left(-1 + \cos\left(\alpha\frac{c_0 t}{R_0}\right)\right)$$

$$\Delta v_i = \frac{e}{c_0 m_i} \frac{\beta}{(R_0)^2} \sin\left(\alpha \frac{r}{R_0}\right) \sin\left(\alpha \frac{c_0 t}{R_0}\right)$$

The obtained expressions imply the applicability of the perturbation theory for the nanometer range bubbles,

$$\frac{\Delta n}{n_0} = \frac{e}{m_i c_0^2} \frac{0.02 \frac{h^2}{e m_e} \sqrt[4]{N} \frac{(3\pi^2)^{\frac{2}{3}}}{12\pi} \left(\frac{4}{3}\pi\right)^{1/3} \alpha^2}{(R_0)^2} \sim 10^{-1},$$

as well as the expression for the speed of sound,

$$c = c_0 + \frac{c_0}{3} \frac{e}{m_i c_0^2} \frac{\beta}{(R_0)^2} \cos\left(\alpha \frac{r}{R_0}\right)\left(-1 + \cos\left(\alpha \frac{c_0 t}{R_0}\right)\right) \quad (20)$$

The formation of shock waves in the class of flows (20) is described in detail in the classical textbook [10]. The time of the shock wave formation in such spherically symmetric system of radius $R_0$ is estimated as

$$t = \frac{R_0}{\pi c_0 \frac{\Delta n}{n_0}} \sim \frac{R_0 m_e m_i c_0^2 (R_0)^2}{c_0 h^2 \sqrt[4]{N}}.$$

For the values of $R_0 \sim 1$ nm, this time is comparable with the time required for sound waves to propagate over a distance of the order of the gas bubble size which is about $t \sim 10^{-14}$ s. Thus, a system of shock waves can be generated in nanometer gas bubbles, which can lead to a significant cumulation of compression.

The possibility of formation of shock waves in a bubble with internal content initially at rest is associated with the existence of an electrostatic potential $\varphi_{0\ electr}^{interect} \sim 0.02 \frac{e N^{0.45} \alpha^2}{4\pi \varepsilon_0 R_0 \left(\frac{4}{3}\pi\right)^{1/3}}$ in the system at the initial moment, and accordingly, with the stored electrostatic energy, which is transformed into the kinetic energy of motion with velocity of the order

$$V \sim \sqrt{\frac{e^2}{\varepsilon_0 m_i R_0}} \sim 10 \; km/s.$$

Note that under conditions of spherical cumulation, the value of the velocity can grow significantly in the central regions of the bubble.

The above analysis demonstrates the formation of shock waves caused by the oscillatory part of the effective potential. Now, we analyze the influence of the barrier potential on the cumulation mechanisms. The electrical charge

$$Q = e n_0 4\pi (R_0)^2 \delta,$$

is distributed at the initial moment over a spherical boundary layer whose thickness $\delta$ is of the order of several interatomic distances The following estimate

$$E \sim \frac{Q^2}{4\pi\varepsilon_0 R_0}.$$

is obvious for the electrostatic energy associated with this charge distribution. This energy is transformed into the kinetic energy of the bubble ions:

$$K = \frac{\frac{4}{3}\pi(R_0)^3 m_i n_0 V^2}{2}.$$

From here, the estimate for the ion mass velocity $V$ follows straightforwardly:

$$V \sim \sqrt{\frac{e^2 n_0^{1/3}}{\varepsilon_0 m_i}} \sim 10 \ km/s.$$

This analysis clarifies that a set of shock waves forms due to the presence of the effective potential in the system during the process of relaxation. The possibility of the formation of shock waves in a bubble with interior initially at rest is associated with the existence of an electrostatic potential in the system at the initial moment, and accordingly, the stored electrostatic energy, which transforms into the kinetic energy of motion.

2.2 The adiabatic compression process

In the second problem, the compression of the gas bubble with the boundary evolution $R(t) = R_0 - U_0 t$ is considered. The adiabatic compression mode is assumed. This implies the fulfillment of the following conditions: (i) The boundary velocity $U_0$ is much less than the speed of sound in the gas bubble, and (ii) the characteristic compression time $t$ is much longer than the time $R_0/C_0$ of propagation of the sound wave through the bubble. In contrast to the standard mode, we take into account the large scale effect.

There exists an exact solution that describes the regime of classical adiabatic compression of this system in the absence of an oscillatory potential:

$$\rho_i^0(t) = \frac{M_{gas}}{\frac{4}{3}\pi(R(t))^3}, \quad n_0(t) = \frac{\rho_i^0(t)}{m_i},$$

$$T_i^0(t) = \left(\rho_i^0(t)\right)^{\frac{2}{3}} \frac{T_i^0(0)}{\left(\rho_i^0(0)\right)^{\frac{2}{3}}},$$

$$u_i(r, t) = -U_0 \frac{r}{R(t)}.$$

The presence of the oscillatory potential is taken into account in the framework of the first order perturbation theory approximation:

$$n(t) = n_0(t) + \Delta n(r, t),$$

$$v_i = u_i(r,t) + \Delta v_i(r,t),$$

$$T_i = T_i^0(t) + \Delta T_i(r,t),$$

$$P_e = \frac{(3\pi^2)^{\frac{2}{3}}}{5}\frac{h^2}{m_e}n^{\frac{5}{3}} = \frac{(3\pi^2)^{\frac{2}{3}}}{5}\frac{h^2}{m_e}n_0^{\frac{5}{3}}\left(1 + \frac{5}{3}\frac{\Delta n}{n_0}\right).$$

Within this approximation, the equations of motion of ions become:

$$0 = -\frac{1}{m_e n_0(t)}\frac{(3\pi^2)^{\frac{2}{3}}}{3}\frac{h^2}{m_e}n_0^{\frac{2}{3}}\frac{\partial \Delta n(r,t)}{\partial r} - \frac{e}{m_e}\frac{\partial}{\partial r}(U_{eff}) - \frac{e}{m_e}\frac{\partial}{\partial r}\varphi \quad (21)$$

$$\frac{\partial \Delta n(r,t)}{\partial t} + \frac{\partial}{\partial r}\left(n_0(t)\Delta v_i(r,t) + \Delta n(r,t)u_i(r,t)\right) + \frac{2}{r}\left(n_0(t)\Delta v_i(r,t) + \Delta n(r,t)u_i(r,t)\right) = 0$$

(22)

$$\frac{\partial \Delta v_i}{\partial t} + u_i(r,t)\frac{\partial \Delta v_i}{\partial r} = -\Delta v_i\frac{\partial u_i(r,t)}{\partial r} - \frac{k}{m}\frac{\partial \Delta T_i(r,t)}{\partial r} - \frac{kT_i^0(t)}{mn_0(t)}\frac{\partial \Delta n(r,t)}{\partial r} - \frac{1}{mn_0(t)}\frac{(3\pi^2)^{\frac{2}{3}}}{3}\frac{h^2}{m_e}n_0^{\frac{2}{3}}\frac{\partial \Delta n(r,t)}{\partial r} - \frac{e}{m_i}\frac{\partial}{\partial r}(U_{eff})$$

(23)

$$\frac{3k}{2m}\left(\frac{\partial \Delta T_i}{\partial t} + u_i(r,t)\frac{\partial \Delta T_i}{\partial r}\right) - \frac{kT_i^0(t)}{mn_0(t)}\left(\frac{\partial \Delta n}{\partial t} + u_i(r,t)\frac{\partial \Delta n}{\partial r}\right) = \frac{k}{mn_0}\frac{\partial n_0(t)}{\partial t}\left(-T_i^0(t)\frac{\Delta n}{n_0} + \Delta T\right) - \frac{u_i e}{m_i}\frac{\partial}{\partial r}\varphi$$

(24)

In the adiabatic compression mode, the profile $\Delta n(r,t)$ will adjust to the profile $U_{eff}$. Therefore, we attempt to look for a solution of the above system of equations in the form:

$$\Delta n(r,t) = A(t)\cos\left(\alpha\frac{r}{R(t)}\right) + D(t)r\sin\left(\alpha\frac{r}{R(t)}\right) \quad \ldots(25)$$

$$\Delta T(r,t) = B(t)\cos\left(\alpha\frac{r}{R(t)}\right) + E(t)r\sin\left(\alpha\frac{r}{R(t)}\right) \quad \ldots(26)$$

$$\Delta v_i(r,t) = C(t)r\cos\left(\alpha\frac{r}{R(t)}\right) + F(t)r^2\sin\left(\alpha\frac{r}{R(t)}\right) \quad \ldots(27)$$

Within the ansatz (25)-(27) the equations (21)-(24) lead to the identical relations that are presented below.

Collecting the terms with $\cos\left(\alpha\frac{r}{R(t)}\right)$ we obtain:

$$\frac{\partial A(t)}{\partial t} + 3\left(n_0(t)C(t) - A(t)\frac{U_0}{R(t)}\right) = 0 \quad ,$$

$$-\frac{k}{m}B(t) - \frac{kT_i^0(t)}{mn_0(t)}A(t) - \frac{1}{mn_0(t)}\frac{(3\pi^2)^{\frac{2}{3}}}{3}\frac{h^2}{m_e}n_0^{\frac{2}{3}}A(t) - \frac{e}{m_i}\frac{\beta}{(R(t))^2} = 0,$$

$$\frac{3k}{2m}\left(\frac{\partial B(t)}{\partial t}\right)_i - \frac{kT_i^0(t)}{mn_0(t)}\left(\frac{\partial A(t)}{\partial t}\right) = \frac{k}{mn_0}\frac{\partial n_0(t)}{\partial t}\left(-T_i^0(t)\frac{A(t)}{n_0} + B(t)\right).$$

It follows from here:

$$C(t) = 0, \quad B(t) \sim R(t)A(t) \sim \frac{1}{(R(t))^2}, \quad A(t) = \frac{0.06\alpha^2 N^{0.45}}{4\pi R_0^3}.$$

We estimate the applicability of the perturbation theory as follows:

$$\frac{\Delta n}{n_0} \sim \frac{0.02\alpha^2}{N^{0.55}} \sim 0.05 \ll 1.$$

It is clear that applicability condition holds with a large margin.

Similarly, collecting the terms with $sin\left(\alpha\frac{r}{R(t)}\right)$ one obtains

$$\frac{\partial D}{\partial t} - A(t)\alpha\frac{U_0}{(R(t))^2} + 4n_0(t)F(t) - 4D(t)\frac{U_0}{R(t)} = 0,$$

$$-kE(t) - \frac{kT_i^0(t)}{n_0(t)}D(t) - \frac{1}{n_0(t)}\frac{(3\pi^2)^{\frac{2}{3}}}{3}\frac{h^2}{m_e}n_0^{\frac{2}{3}}D(t) = 0.,$$

$$3k\frac{\partial E(t)}{\partial t} - \frac{kT_i^0(t)}{n_0(t)}\left(\frac{\partial D(t)}{\partial t}\right) = \frac{k}{n_0}\frac{\partial n_0(t)}{\partial t}\left(-T_i^0(t)\frac{D(t)}{n_0} + E(t)\right),$$

$$+ \frac{U_0}{R(t)}\left(\alpha\frac{\beta e}{(R(t))^3} + \frac{1}{n_0(t)}\frac{(3\pi^2)^{\frac{2}{3}}}{3}\frac{h^2}{m_e}n_0^{\frac{2}{3}}A(t)\frac{\alpha}{R(t)}\right).$$

Hence, we get $F(t) = \frac{0.06\alpha^3}{12R_0^2}\frac{U_0}{N^{0.55}}$

Now let us estimate the time of the shock wave formation by considering the trajectory of the Lagrangian particle $r(t, r_0)$ with the initial coordinate. The equation of motion has the following form:

$$\frac{dr(t,r_0)}{dt} = u_i\left(r(t,r_0), t\right) + \Delta v_i\left(r(t,r_0), t\right).$$

Applying the method of successive approximations, we obtain

$$r(t, r_0) = r^{(0)}(t, r_0) + r^{(1)}(t, r_0),$$

$$r(t,r_0) = (R_0(0) - U_0 t)\frac{r_0}{R_0(0)} + \frac{0.06\alpha^3}{12N^{0.55}}\frac{r_0^2}{(R_0(0))^2}\cos\left(\alpha\frac{r_0}{R_0(0)}\right)(R_0(0) - U_0 t)\ln\left[\frac{(R_0(0)-U_0 t)}{R_0(0)}\right].$$

The condition for the occurrence of a shock wave is the intersection of the trajectories of Lagrangian particles:

$$r(t,r_0) = r(t, r_0 + \Delta) \text{ for } \Delta \neq 0.$$

Hence, taking into account the smallness of $\Delta \ll r_0$, we have

$$\Delta = \frac{0.01\alpha^3}{N^{0.55}}\frac{r_0^2}{(R_0(0))}\cos\left(\alpha\frac{r_0}{R_0(0)}\right)\cos\left(\alpha\frac{\Delta}{R_0(0)}\right)\ln\left[\frac{(R_0(0)-U_0 t)}{R_0(0)}\right]$$

It follows that, under the requirement $\Delta = \frac{R_0(0)}{10}$ for the formation of a shock wave in the bubble volume $\frac{r_0}{(R_0(0))} \sim 0.5$, shock waves can only be formed under the unrealistic condition $\frac{(R_0(0)-U_0 t)}{R_0(0)} \sim 10^{-2}$. That is, during adiabatic compression of a relaxed system, shock waves are not formed.

Thus, quantum shell effects in hydrodynamic processes manifest themselves only in the processes of system relaxation.

2.3 Numerical simulation results

In this section, we present results of numerical simulation that confirm the existence of a non-classical mechanism of the compression of a submicron gas bubble. Suppose cold gas fills a spherical cavity, surrounded by dense medium, e.g. diamond. The diamond is a good insulator, so its border acts as quantum well potential with high amplitude. The effect described above emerges. When the system is compressed by external pressure, the effective force due to the quantum nature of the degenerate electronic component appends to gradients of pressure.

In simulations we consider the compression of hydrogen gas inside a cavity with $R$=1 nm. The boundaries of the cavity converge towards the center with constant velocity. Initially gas has uniform density ρ=5 g/cm3, so the system is taken unrelaxed to new external force. Initial temperature $T$=0.1 eV. Simulations are carried out in one-temperature single-fluid approximation (the dynamics of ionic components is simulated explicitly). Details of the numerical code used is given in Appendix 3 and [11].Such set up goals to demonstrate the maximum possible impact by external force, which allows to estimate the prospect of the effect.

Two calculations of submicron bubble compression are performed. The ionic system at the initial moment is taken to be unrelaxed in these calculations.

The first calculation is carried out with the effective potential (13), where $U_{osc}(r)$ and $U_{bar}(r)$ are the oscillator and barrier potentials, which are modeled by the following expressions:

$$U_{bar} = -\frac{eV_0}{1+\exp[(R_0 - r)/\delta]}$$

$$U_{osc} = C\left(\frac{\bar{n}}{10^{30}}\right)^{1/3} \frac{R_0(0)}{R_0(t)} f(r, R_0(t))$$

In the second problem, we calculate the compression of the ion gas without taking into account the influence of the effective field $U_{eff}$=0. Comparison of the obtained results allows one to evaluate the magnitude of the effect.

The numerical calculations are carried out in dimensionless variables. The corresponding scales and transformation of the main hydrodynamic variables to the dimensionless form are described in Appendix 3.
The following parameters are chosen for numerical calculations: the bubble radius is 1 nm; the gas density ρ=5 g/cm³ ; the velocity of the bubble bound $u_w$=-2 km/sec is adopted; time interval $t_c$=0.45, which corresponds to 90% gas compression; the barrier and oscillation potentials parameters are fixed at the values $\delta$= 0.1$R_0$ and C=10eV, $V_0$=1V.

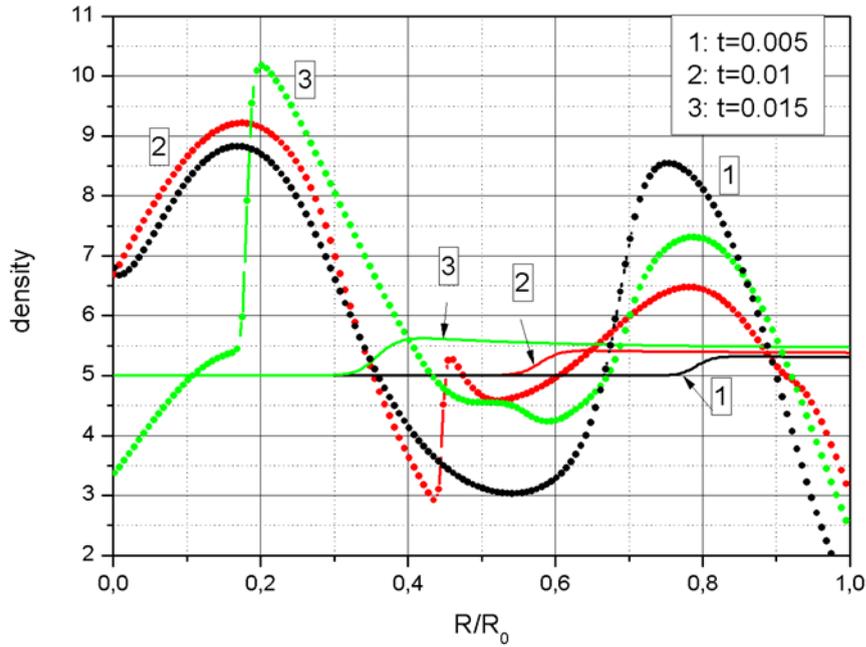

Fig.1. Radial dependence of density (g/cm³) for successive times $t$=0.005, 0.01, 0.015 (early stage of compression). The dotted line shows results of calculations with account for the $U_{eff}$, the solid line is related to calculations without the $U_{eff}$,

The results of numerical simulation are shown in Figs. 1-4. Radial dependence of density, velocity, pressure, and temperature are plotted for three successive time points corresponding to the initial stage of compression (about 3%).

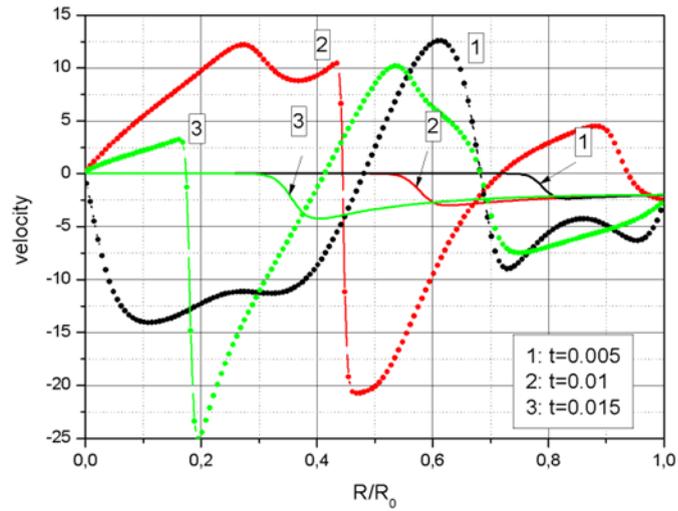

Fig. 2. Radial dependence of velocity (km/s) for successive times t=0.005, 0.01, 0.015 (early stage of compression). The dotted line shows results of calculations with account for the $U_{eff}$, the solid line is related to calculations without the $U_{eff}$,

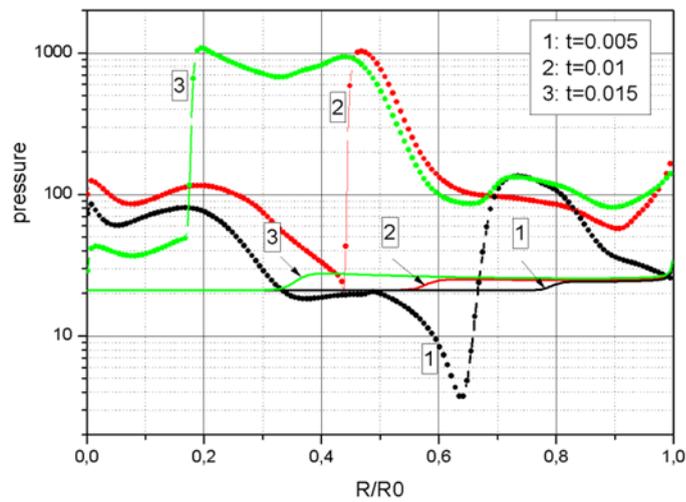

Fig. 3. Radial dependence of pressure for successive times t=0.005, 0.01, 0.015 (early stage of compression). The dotted line shows results of calculations with account for the $U_{eff}$, the solid line is related to calculations without the $U_{eff}$,

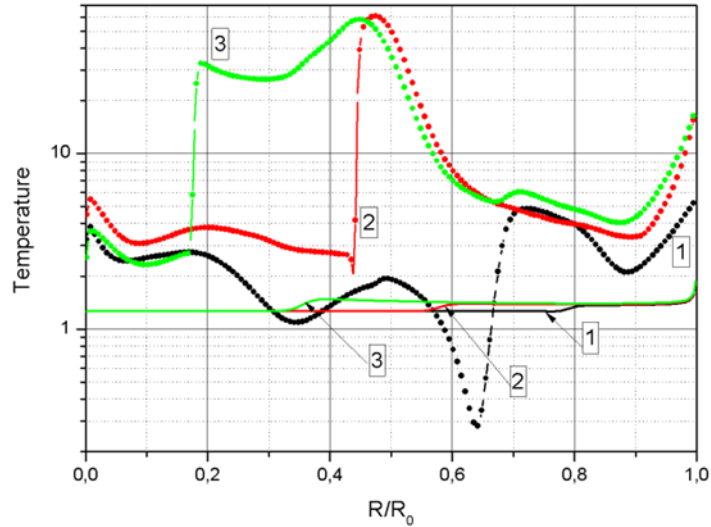

Fig. 4. Radial dependence of temperature ($10^3$ K) for successive times t=0.005, 0.01, 0.015 (early stage of compression). The dotted line shows results of calculations with account for the $U_{eff}$, the solid line is related to calculations without the $U_{eff}$.

Let us compare the results of calculations without an external electric field (solid line) and with presence of an effective field (dashed line). In the absence of the electric field, compression evolves in accordance with the theory of classical gas dynamics, with the formation of a shock wave moving towards the center (shown by arrows in the figures). A qualitatively different picture arises if the effective potential is taken into account. The motion of the wall brings the system out of equilibrium, and the resulting field strongly affects the motion of the ions. A compression wave appears by the time t=0.005, and the amplitude of the wave is several orders of magnitude higher than the amplitude of the shock in a purely gas-dynamic calculation. At subsequent moments, this wave overshoots forming a shock wave with a relative amplitude of about 1000. The velocity in the wave reaches a value of 20 km/sec.

Reaching the center, the wave causes a focusing of energy with a sharp rise in pressure and temperature in the center. This is seen in Figs. 5 - 7, which show the time dependence of the density, pressure and temperature at the center of the spherical bubble.

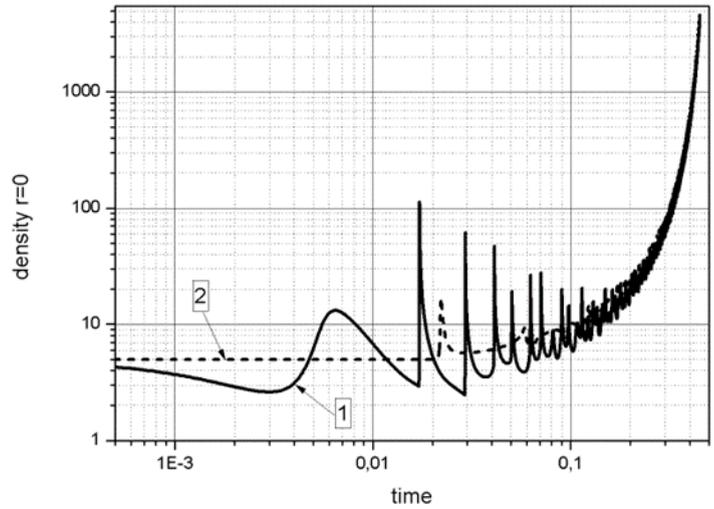

Fig. 5. Time dependence of density (g/cm$^3$) at the center of a spherical bubble. The solid line shows results of calculations with account for the $U_{eff}$, the dotted line is related to calculations without the $U_{eff}$,

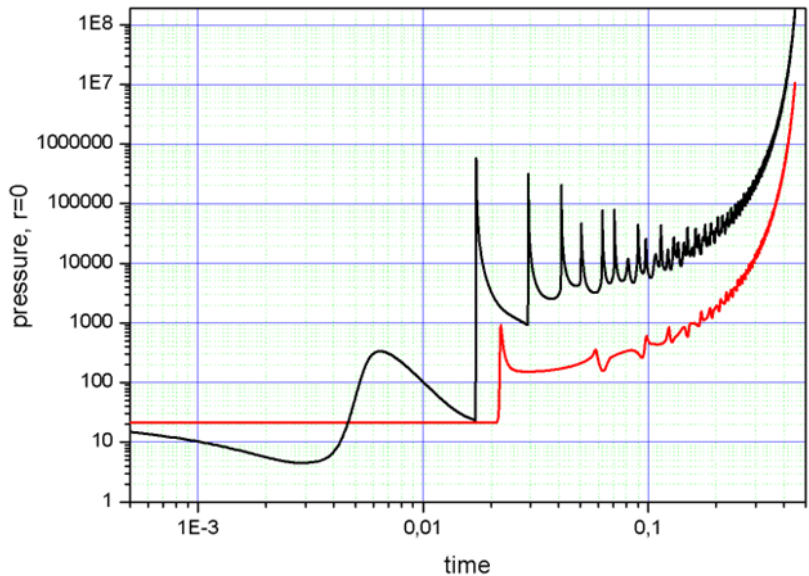

Fig. 6. Time dependence of pressure at the center of the spherical bubble. The black line shows results of calculations with account for the $U_{eff}$, the red line is related to calculations without the $U_{eff}$.

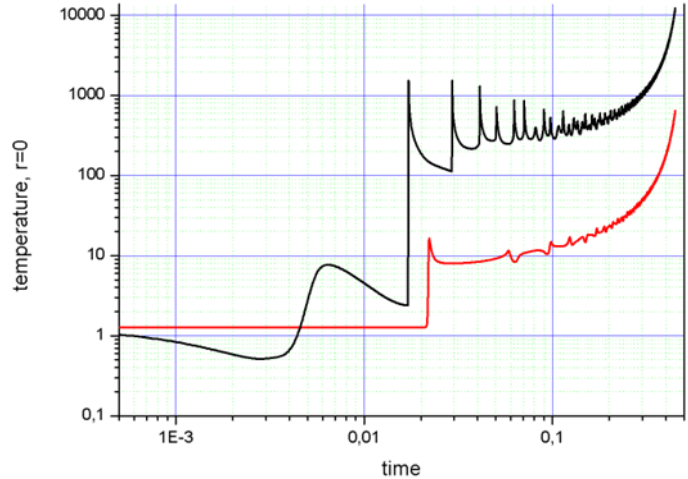

Fig.. 7. Time dependence of temperature ($10^3$ K) at the center of the spherical bubble. The black line shows results of calculations with account for the $U_{eff}$, the red line is related to calculations without the $U_{eff}$.

Ordinary gas-dynamic calculations indicate the pressure increase by about 2 orders of magnitude when the shock wave focuses. Taking into account the effective field enhances this effect by almost 1000 times (Fig. 6). Similar effects emerge on the temperature graph. The temperature at the center increases by 2 orders of magnitude, reaching a value of ~1000 ($10^3$ K), at the moment of focusing of the shock wave which formed as a result of the compression wave overshooting due to the field.

The process of compression of the electron-ion system is accompanied by a series of multiple reflections of shock waves from the surface of the spherical bubble and subsequent focusing at the center.

Thus, comparison of the results of calculations reveals that taking into account the quantum effect of a large spatial scale in the distribution of electrons qualitatively changes the nature of cumulative processes.

Due to the large-scale electric field caused by quantum shell effects, the compression process is characterized by the formation of multiple shock waves. The values of gas temperature and pressure reached during compression turn out to be approximately two orders of magnitude higher as compared with the ordinary compression regime.

# Conclusion

A new mechanism of hydrodynamic cumulation caused by quantum shell effects in a mesoscopic system consisting of a gas of degenerate electrons and a gas of classical ions has been invented and studied extensively.

The direct formulation, which implies self-consistent calculation of the gas-dynamic motion of ions and solving the quantum mechanical problem for determining the spatial distribution of the mesoscopic number of electrons, encounters significant computational complexity. Therefore, the simplified approach was elaborated. The main ingredients of such an approach are: hydrodynamic description of the ionic and electronic subsystems and introduction of an external effective field acting on the electronic subsystem. This effective field allows one to take into account quantum effects in a simplified manner. The obtained results demonstrate validity and efficacy of the developed approach for the analysis of the considered class of problems.

The developed approach demonstrates that due to the presence of a large-scale electric field caused by quantum shell effects the process of gas bubble compression is accompanied by the formation of multiple shock waves. This drastically changes the character of cumulative processes. The values of temperature and pressure in the central region increase by several orders of magnitude compared to the conventional compression regime.

The experimental study of the described effect is among the key future tasks. The temperature increasing in the bubble due to this effect could lead to launching thermonuclear reactions. Therefore, the obtained fusion neutrons can be used for diagnostics. The preliminary simplified calculations showed that cumulation in a nanoscale diamond target with DT-ice central cavity could generate some neutrons. In this case, the energy spent on the generation of one neutron is comparable to the NIF record one [4]. If there is no effect, considered by us, then launching a thermonuclear fusion in such a system is impossible.

A difficult issue is the choice of a source for irradiating such a target. The direct laser irradiation requires the use of an extremely short wavelength laser due to the diffraction limit. On the other hand, the hohlraum experiment could break the implosion symmetry that is crucial for the considered effect. Moreover, the X-ray radiation could preheat the DT-ice cavity and hence violate the necessary conditions.

Nevertheless, the main problem of such an experiment is that one target generates a number of neutrons that is well below the detectability threshold of the neutron detectors on the main laser facilities. The intuitive desire to increase the number of targets and place them in the hohlraum faces a number of serious technical problems. The symmetry and synchronicity issues depend on how large a number of nano-targets would be placed in a hohlraum. The way it is done currently is by embedding them in a sub-critical-density plastic or carbon foam. It has been shown, however, both theoretically [28] and experimentally [29] that the laser generated heat front propagation velocity in such foams is microstructure dependent. The development of the experiment idea will be the subject of our following studies.

## Acknowlegments

The authors are grateful to Y.E. Lozovik for very helpful comments.

# Appendix 1. The Effective Potential

The effective potential is determined from the relation

$$\frac{1}{n_e^{free}}\frac{\partial p_e(n_e^{free})}{\partial r} = -e\frac{\partial}{\partial r}(U_{eff}(r)) \qquad (A1.1)$$

which is a consequence of the Euler equation of motion for the electron fluid and the spherical symmetry of the problem.

The electronic subsystem is considered within the Fermi gas model. Accordingly the pressure and density of the electronic subsystem are related in the following way:

$$p_e = \frac{(3\pi^2)^{2/3}}{5}\left(\frac{h^2}{m}\right)n_e^{5/3} \qquad (A1.2)$$

In further calculations, it is convenient to use the dimensionless concentration $\tilde{n}_e$ defined by the relation

$$\tilde{n}_e = 4\pi R_0^3 n_e \qquad (A1.3)$$

$$\int_0^1 \tilde{n}_e \zeta^2 d\zeta = N_e \qquad (A1.4)$$

where $\zeta = r/R_0$ is the dimensionless radial coordinate, and $N_e$ is the number of electrons. The equation

$$\frac{(3\pi)^{2/3}}{2^{5/3}}\frac{h^2}{mR_0^2}\frac{d\tilde{n}_e^{2/3}}{d\zeta} = \frac{dU_{eff}}{d\zeta} \quad, \qquad (A1.5)$$

follows directly from relations (A1.1)-(A1.4). This equation has the obvious solution

$$U_{eff} = \frac{(3\pi)^{2/3}}{2^{5/3}}\frac{h^2}{mR_0^2}\tilde{n}_e^{2/3} + const \quad . \quad (A1.6)$$

Due to the smallness of density fluctuations $\delta\tilde{n}_e$ as compared with the mean value $\tilde{n}_e^0$, formula (A1.6) can be simplified

$$U_{eff} = \frac{(3\pi^2)^{2/3}}{2^{5/3}}\frac{h^2}{mR_0^2}\left(\tilde{n}_e^0 + \delta\tilde{n}_e\right)^{2/3} \approx \frac{(3\pi^2)^{2/3}}{2^{5/3}}\frac{h^2}{mR_0^2}\left(\left(\tilde{n}_e^0\right)^{2/3} + \frac{2}{3}\frac{\delta\tilde{n}_e}{\left(\tilde{n}_e^0\right)^{1/3}}\right). \quad (A1.7)$$

Approximation (A1.7) loses its accuracy and becomes inapplicable near the wall ($r \sim R_0$, or $\zeta \sim 1$), which is due to vanishing of the electron density on the wall in accordance with the accepted boundary condition of the impenetrability of the wall in the quantum mechanical problem.

Figures A1.1-A1.3 show examples of non-averaged dependences $\tilde{u} = \tilde{n}_e^{2/3} - \left(\tilde{n}_e^0\right)^{2/3}$ for different values of the parameter $k_f R_0 \sim N^{1/3}$. The electron density is determined by means of direct summation over all the occupied states of electrons in the spherical potential well with infinite walls. Another technique based on Green's function approach is also suitable for numerical calculation of the electron density. The method of constructing the Green's function based on two linearly independent solutions occurs rather conveniently for this purpose. As it is shown in Ref.

[2] the results, obtained within these two ways coincide with high accuracy. In itself the extent to which these results are consistent can be considered as an indicator of numerical calculations accuracy.

With an appropriate choice of the constant in (A1.7), the quantity $\tilde{u}$ is proportional to $U_{eff}$ and can be considered as a dimensionless effective potential. Since the gas-dynamic calculation implies the use of averaged values over a large number of particles, the direct use of the value (A1.7) for such purposes is not correct. The density determined in the framework of quantum mechanics must be preliminarily averaged in a suitable way and only then it should be included in gas-dynamic calculations. The averaging procedure looks as follows. Due to spherical symmetry the averaging should be performed over radial variable. For these purposes, we use averaging with the Gaussian weight function: $\exp\left(-\frac{(r-r')^2}{a^2}\right)$. The mean values of a certain quantity has the form

$$\bar{f}(r) = \frac{1}{C(r)} \int_0^{R_0} \exp\left(-(r-r')/a^2\right) f(r') dr' \quad , \quad (A1.8)$$

where $\frac{1}{C(r)}$ is the normalization factor,

$$C(r) = \int_0^{R_0} \exp\left(-(r-r')/a^2\right) dr' \quad .$$

Due to the finite averaging interval $[0, R_0]$ the factor $C(r)$ is coordinate dependent which should be taken into account when calculating derivatives of $\bar{f}(r)$.

Results of averaging procedure (A1.8) with different values of smearing parameter $a = 300\ h_r$, $600\ h_r$, $800 h_r$ and $1000\ h_r$ ($h_r$ is the increment of uniform radial grid, the total number of grid nodes is equal 10000) are presented in Figures A1.4-A1.6. It can be seen that for $a = 600$, convergence in the averaged potential is achieved. This value of $a$ was adopted below.

Figures A1.7-A1.11 demonstrate the results of averaging of the function $\tilde{u} = \tilde{n}_e^{2/3} - \left(\tilde{n}_e^0\right)^{2/3}$ for $k_f R_0 \sim (28.0\text{-}33.0)$. This function differs from the effective potential by a constant factor only. The obtained results allow us to note the following patterns of behavior of $U_{eff}$. The shape of the effective potential depends on the number of particles and has from 1 to 3 bumps. There is some periodicity of the potential shape (see Figure A1.11) when changing the number of particles. The characteristic period is approximately equal to $\Delta k_f R_0 \sim$ (1.3-1.4) in this dependence. The amplitude of the effective potential $U_{eff}$ is proportional to $\frac{1}{R_0^2}$ and weakly depends on $N$. The behavior is similar for other values of $k_f R_0$ (see Figure A1.12).

The electrostatic potential of the electronic subsystem is calculated in the usual way

$$U_c = \int_V \frac{e}{4\pi\varepsilon_0} \frac{1}{|r-r'|} n_e(r') d^3 r' \quad (A1.9)$$

Spherical symmetry simplifies Eq. (A1.9):

$$U_c = \frac{e}{4\pi\varepsilon_0} \frac{1}{R_0} \left\{ \frac{1}{\zeta} \int_0^\zeta \tilde{n}_e(\zeta') \zeta'^2 d\zeta' + \int_\zeta^1 \tilde{n}_e(\zeta') \zeta' d\zeta' \right\} \quad (A1.10)$$

The first term in Eq. (A1.10) is the potential on the surface of the ball of radius *r* and the second one is the potential of a spherical layer on its inner boundary.

We also calculate large-scale spatial oscillations $\varphi_{electr}^{free}$ - the electric potential of an electrically neutral system consisting of non-interacting electrons and a spatially homogeneous ball of ions. This value allows one to estimate the shape and magnitude of the electric force acting on the ionic system.

It should be emphasized that the electrostatic potential defined by Eqs. (A1.9), (A1.10) does not directly connected with effective potential (A1.1)

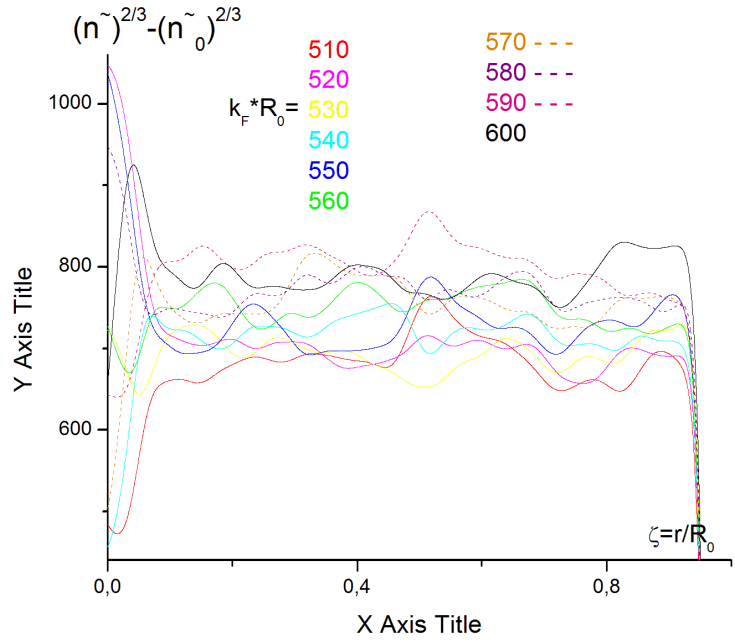

Figure A1.1 Radial dependence of the effective potential for different values of $k_F R_0$.

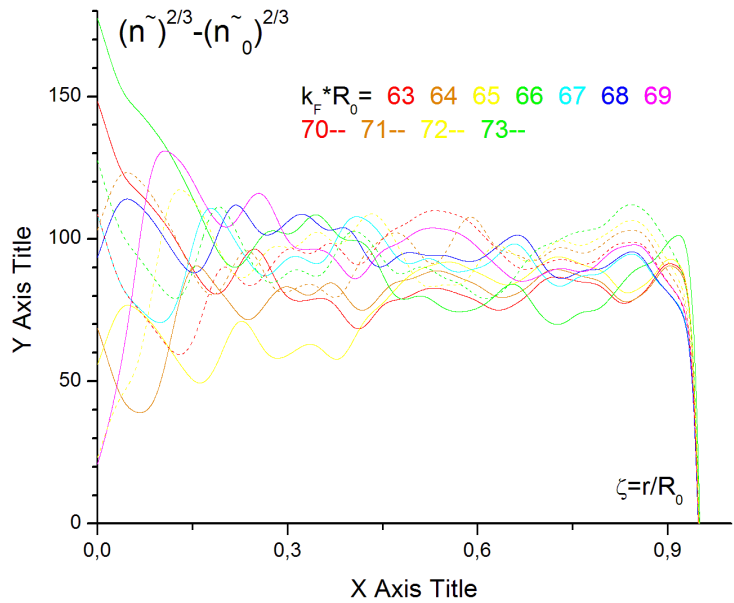

Figure A1.2 Radial dependence of the effective potential for different values of $k_F R_0$.

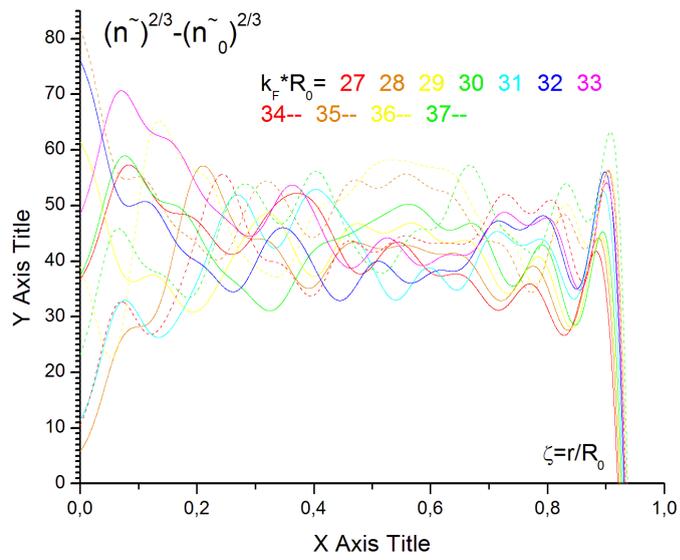

Figure A1.3 Radial dependence of the effective potential for different values of $k_F R_0$.

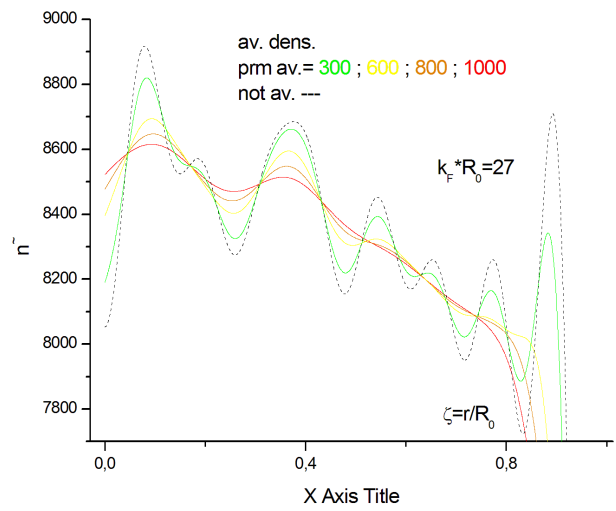

Figure A1.4 Radial dependence of the electron density and averaged ones obtained with different values of smearing parameter; $k_F R_0 = 27$.

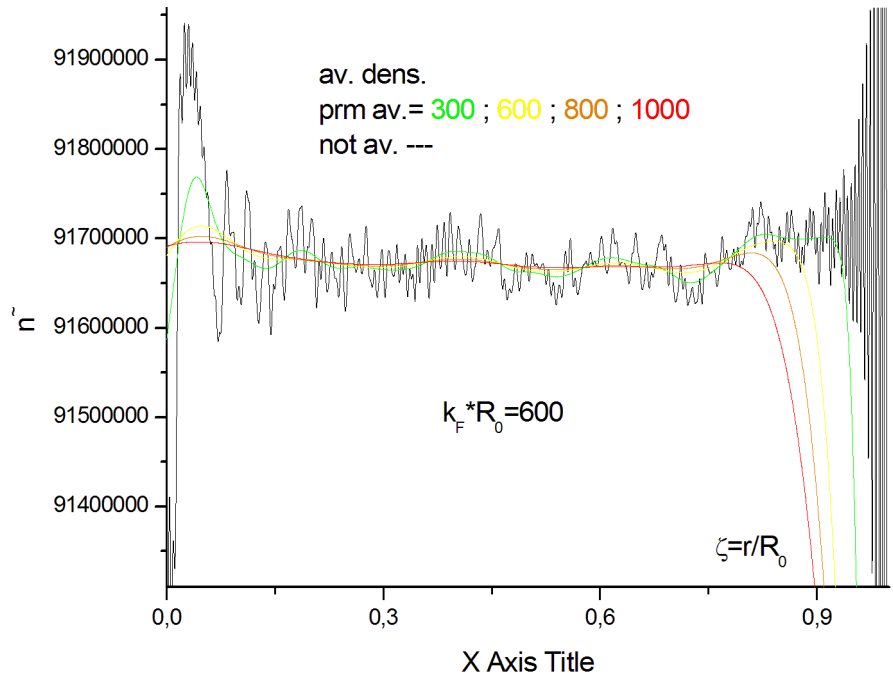

Figure A1.5 Radial dependence of the electron density and averaged ones obtained with different values of smearing parameter; $k_F R_0 = 600$

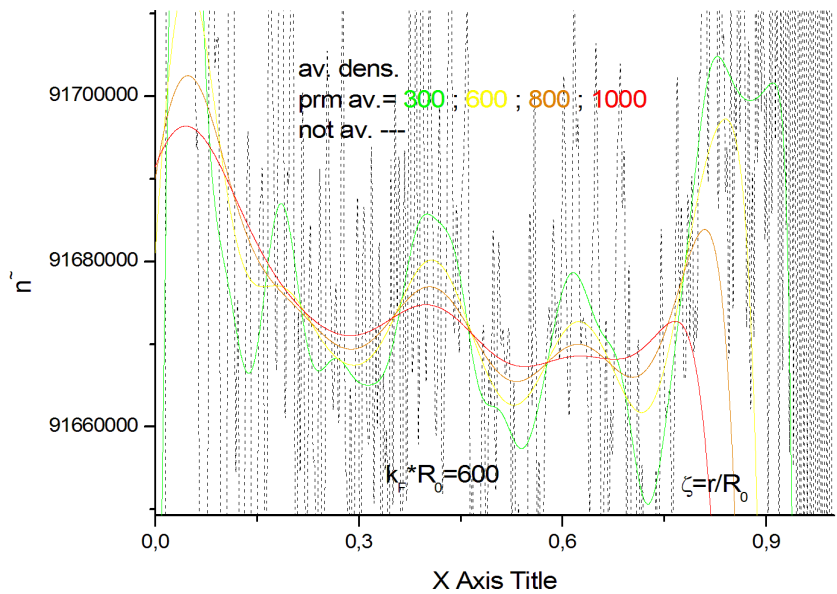

Figure A1.6 The same as in previous Fig. A1.5, but with enlarged scale along vertical axis

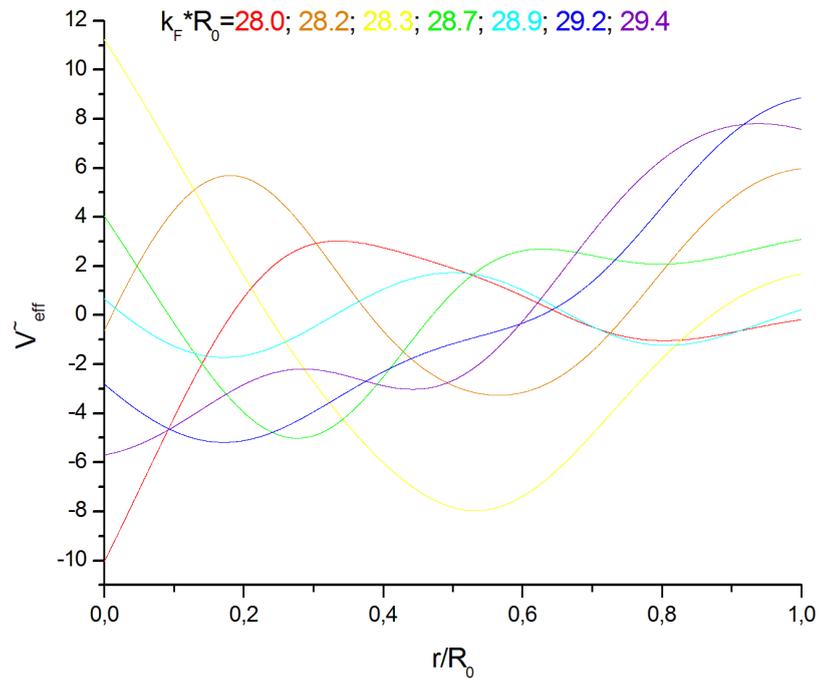

Figure A1.7 Radial dependence of the oscillation potential for several values of $k_F R_0$.

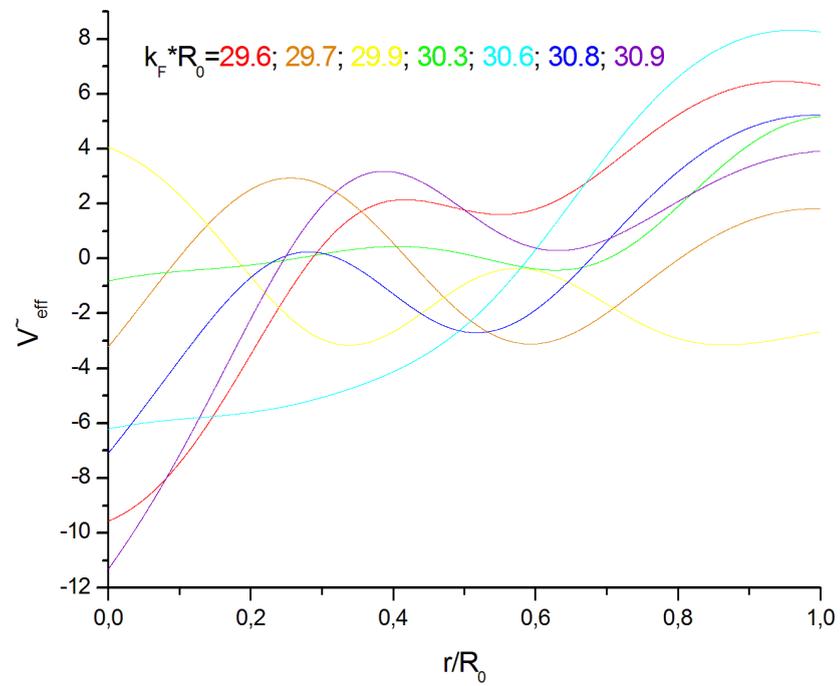

Figure A1.8 Radial dependence of the oscillation potential for several values of $k_F R_0$.

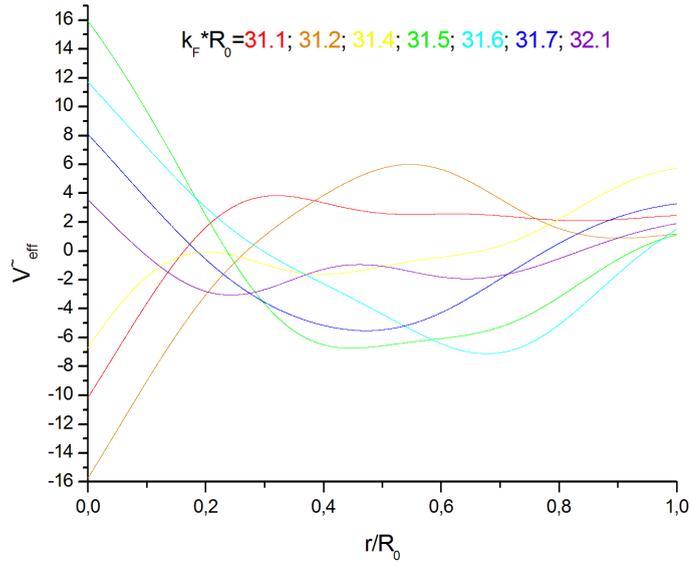

Figure A1.9 Radial dependence of the oscillation potential for several values of $k_F R_0$.

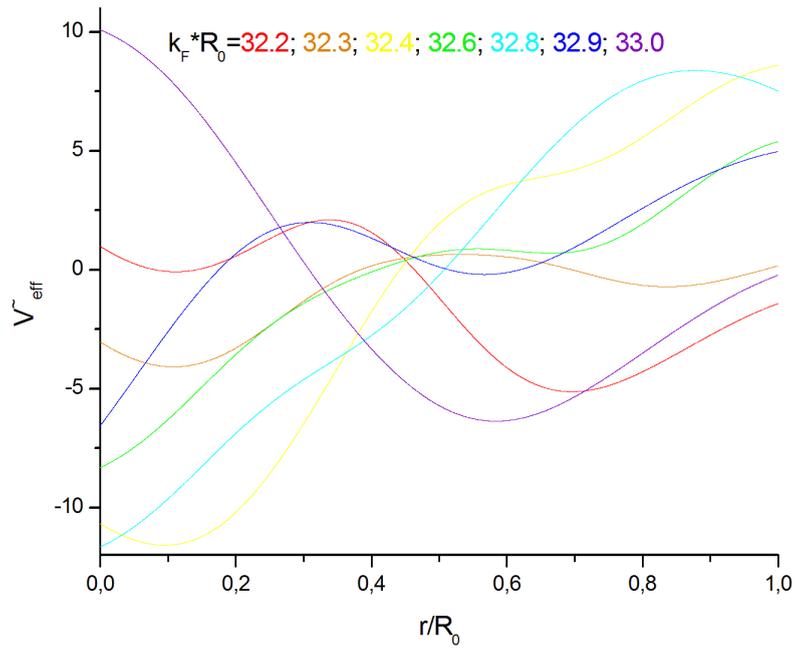

Figure A1.10 Radial dependence of the oscillation potential for several values of $k_F R_0$.

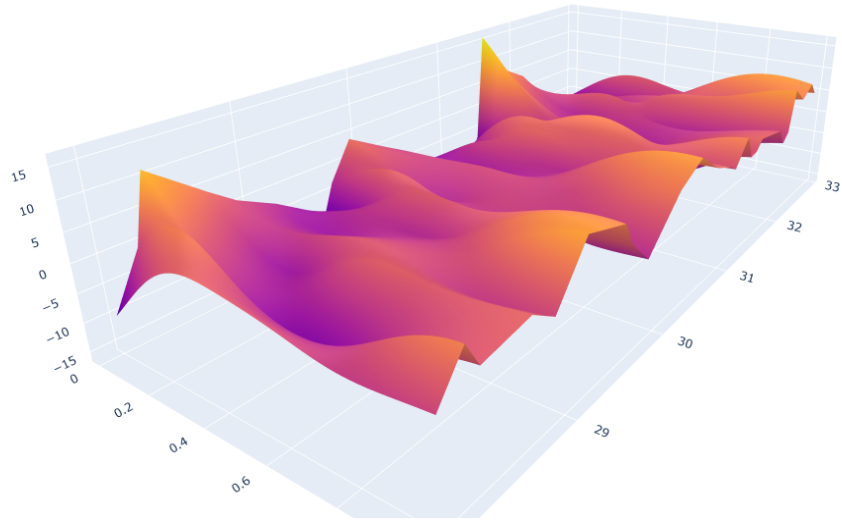

Figure A1.11 3D plot of the dimensionless effective potential as a function of the radial coordinate and the parameter $(k_F R_0)$. Quasi-periodicity appears in the $(k_F R_0)$ dependence.

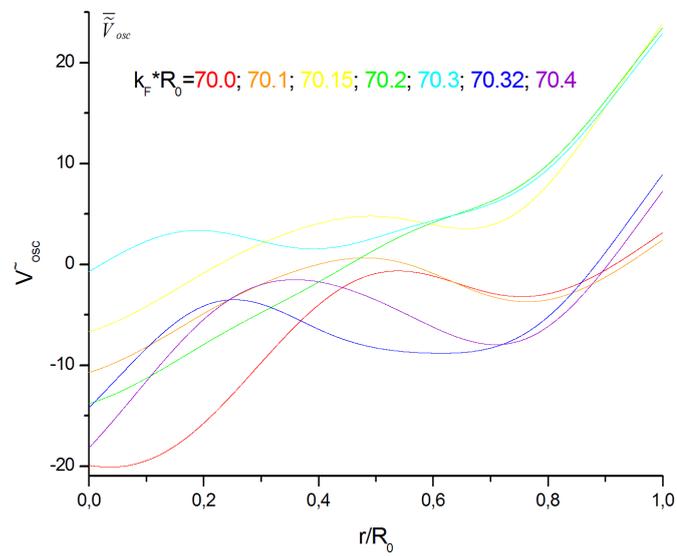

Figure A1.12 Radial dependence of the oscillation potential for several values of $k_F R_0$.

The potential (A1.10) should be supplemented with the potential created by ions. Within the approximate model, a homogeneous ion background is considered, and accordingly its potential is that of a uniformly charged ball. In this case, the following circumstance should be taken into account. The zero boundary condition on the wall of the cavity in the quantum mechanical problem causes vanishing of the electron density on the wall. This circumstance should be taken into account in the model ionic density. Namely, the ionic density is modeled by a uniformly charged ball with a slightly smaller radius than $R_0$. The reference point for choosing the ion distribution radius is zeroing of the total potential of electrons and ions on the wall (i.e., at r=$R_0$).

Averaging, similar to that applied to $U_{eff}$, should also be applied to $U_C$. It should be noted that the radial dependence of the unaveraged potential $U_C$ differs significantly from $U_{eff}$. This is due to $U_{eff}$, is proportional to $\tilde{n}_e^{\sim 2/3}$, which is a rapidly oscillating function, while $U_C$ is defined through integration of $\tilde{n}_e$ in (A1.10). Integration smears out these high-frequency oscillations. Nevertheless, the uniformity of the approach implies that averaging should be applied to $U_C$ also.

Figures A1.13-A1.15 show the results of averaging the electric potentials $\varphi_{electr}^{free}$ (made dimensionless by the parameter $\frac{e}{4\pi\varepsilon_0}\frac{1}{R_0}$) for different values of the parameters $k_f R_0 \sim N^{1/3}$. It can be seen from Figures A1.13-A1.15, that the behavior of the potential $\varphi_{electr}^{free}$ is in qualitative accordance with the shape of the potential $U_{eff}$.

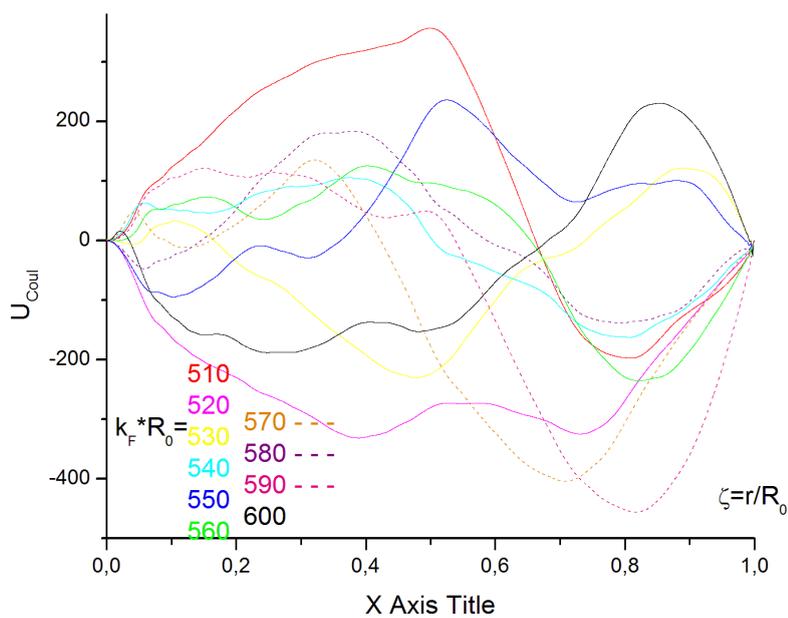

Figure A1.13 Radial dependence of the Coulomb potential for several values of $k_F R_0$

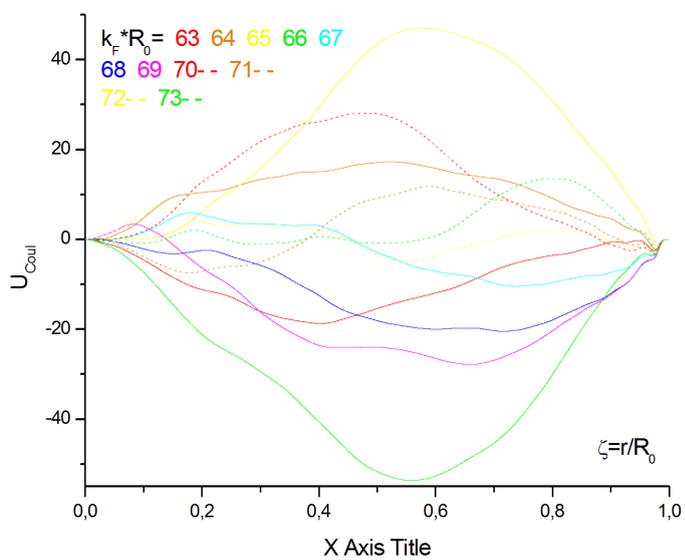

Figure A1.14 Radial dependence of the Coulomb potential for several values of $k_F R_0$

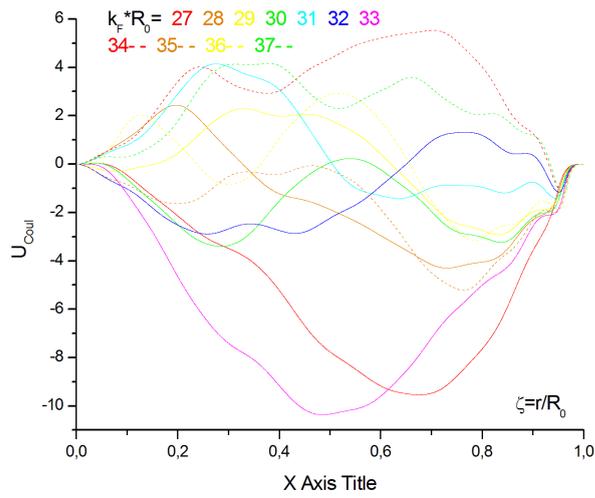

Figure A1.15 Radial dependence of the Coulomb potential for several values of $k_F R_0$

      Based on the calculated values $U_C$ the electron densities for the systems with $k_f R_0$ = 27, 600 are computed by means of the Poisson equation. These results are presented in Figures A1.16-A1.17 (green curve).
Closeness of the results confirms the similarity of the two methods of averaging.

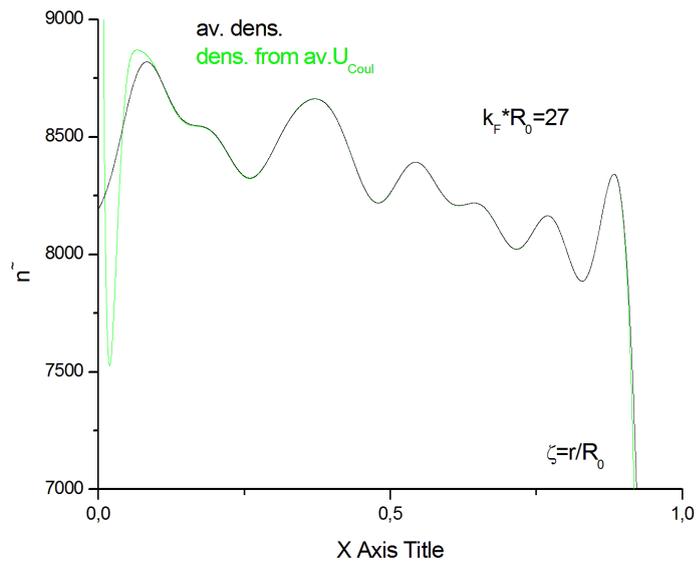

Figure A1.16 Radial dependence of the electron density obtained by means of averaging procedure (A1.8) and that from averaged Coulomb potential; $k_F R_0 = 27$.

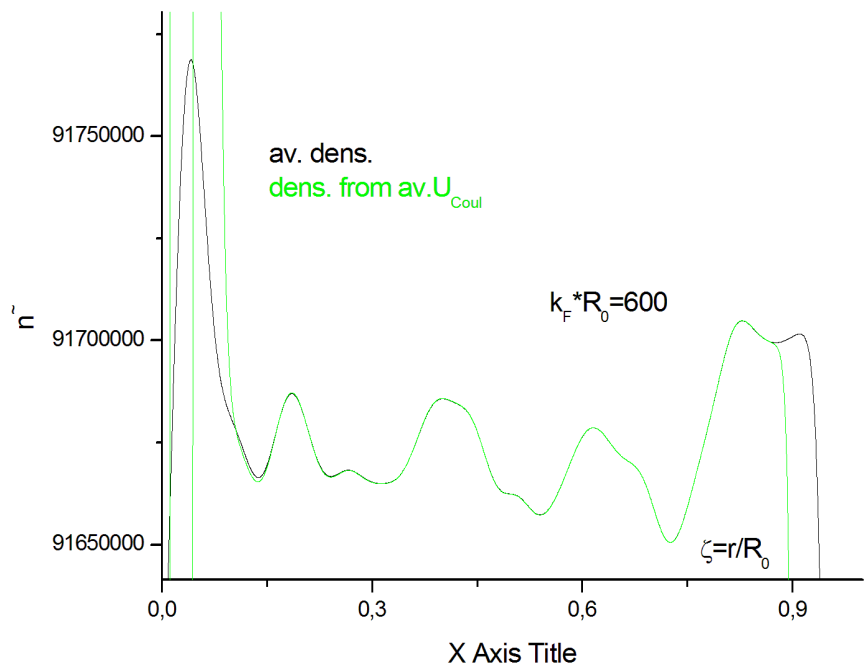

Figure A1.17 Radial dependence of the electron density obtained by means of averaging procedure (A1.8) and that from averaged Coulomb potential; $k_F R_0 = 600$.

# Appendix 2. Model Approximation of the Effective Potential

A characteristic feature of the effective potential $U_{eff}$ is its fluctuating behavior near the mean value, and a sharp drop to zero at the cavity boundary. To convey the abrupt behavior in the near-wall region, the Woods-Saxon function for the barrier potential can be used:

$$f_{WS}(r) = \frac{W}{1+\exp\left((r-R_{WS})/a\right)},$$

where $R_{WS}$ is the distribution radius, $a$ is diffuseness, and the parameter $W$ determines the average value inside the cavity. By varying these three parameters, we achieve the best reproduction of $U_{eff}$, by means of minimizing of the sum of squared deviations of $U_{eff}$ values on the grid from the approximant $f_{WS}$. That is we are looking for:

$$\min\left(\sum_{i=1}^{N}\left(U_{eff}(r_i) - f_{WS}(r_i)\right)^2\right)$$

on the grid $\{r_i\}_{i=1}^{N}$. This part of the effective potential is referred to as barrier potential. The next step in the approximation is to take into account oscillations of $U_{eff}$ near the mean value inside the cavity. Bearing in mind the use of this potential in the gas-dynamic calculation, it should, as noted above, be averaged over high-frequency oscillations in order to reveal large-scale dependencies. Thus, the approximant should reproduce not high-frequency fluctuations, but rather a smoothed behavior. To do this, a rapidly changing quantity $\left(U_{eff}(r_i) - f_{WS}(r_i)\right)$ was preliminarily averaged with a Gaussoid (see (A1.8). The averaged values obtained in this way for the quantity $\overline{\left(U_{eff}(r_i) - f_{WS}(r_i)\right)}$ are much smoother, and the characteristic spatial scale of their change is comparable to the size of the cavity. It is convenient to approximate this dependence by a partial sum of the Fourier series with the retention of 5-10 terms depending on the desired accuracy:

$$\overline{\left(U_{eff}(r_i) - f_{WS}(r_i)\right)} \approx \sum_{m=0}^{M} C_m \cos\cos\left(\pi m \frac{r}{R_0}\right)$$

This part of the effective potential is referred to as oscillation part. It should be noted that averaging and representation by a partial sum of the Fourier series should be carried out only after separation of the barrier potential. The point is that the origin of the barrier potential is related to the boundary condition, which is rather rigid (vanishing of the wave functions on the cavity wall) and is not the effect of the appearance of a large scale in the oscillating part of $U_{eff}$. Table A 2.1 shows the values of the coefficients $\widetilde{C}_m$ that differ from the Fourier coefficients in the above formula by a constant factor $C_m = \frac{(3p)^{2/3}}{2^{5/3}} \frac{h^2}{mR_0^2} \widetilde{C}_m$. The choice of parameter values $k_f$ $R_0$ corresponds to the dependencies shown in Fig. A1.7—A1.10.

Table A 2.1 Expansion coefficients of the Fourier series of the oscillation potential for several values of $k_F R_0$

| $k_F R_0$ | $\tilde{C}_0$ | $\tilde{C}_1$ | $\tilde{C}_2$ | $\tilde{C}_3$ | $\tilde{C}_4$ | $\tilde{C}_5$ |
|---|---|---|---|---|---|---|
| 28.0 | -0.3894 | -0.7383 | -2.8435 | -2.3562 | -0.9652 | -0.8250 |
| 28.2 | 2.9122 | 0.6681 | 3.6358 | -2.1092 | -1.0644 | -0.8042 |
| 28.3 | -3.1570 | 2.9068 | 6.5422 | 0.7503 | 0.4930 | 0.4791 |
| 28.7 | 0.4062 | -2.6594 | 0.8154 | 2.3031 | 1.6555 | 0.2033 |
| 28.9 | -0.3627 | -0.1521 | -1.0935 | -0.0176 | 0.9988 | 0.1052 |
| 29.2 | -0.3176 | -5.7909 | 1.7601 | -0.7042 | 0.8527 | 0.3029 |
| 29.4 | 0.6283 | -5.9594 | 1.9040 | -0.7174 | -1.0703 | -0.0243 |
| 29.6 | 1.7177 | -5.9279 | -1.5893 | -2.1648 | -0.6561 | 0.2338 |
| 29.7 | -0.0855 | 0.4883 | 1.0883 | -2.2426 | -1.0793 | -0.2200 |
| 29.9 | -2.4844 | 1.7689 | 0.7406 | 1.7640 | 1.0054 | -0.2371 |
| 30.3 | 1.5828 | -1.7354 | 1.0854 | -1.1616 | 0.4130 | 0.0061 |
| 30.6 | -0.6460 | -7.3809 | 1.8339 | 0.0736 | -0.3225 | 0.1087 |
| 30.8 | -0.4265 | -3.4219 | 1.1895 | -1.9458 | -1.3187 | -0.2825 |
| 30.9 | 0.1190 | -3.8300 | -2.3894 | -3.1061 | -0.7394 | -0.0401 |
| 31.1 | 2.7634 | -2.1295 | -2.5225 | -2.1742 | -1.4719 | -0.8770 |
| 31.2 | 0.7209 | -4.6133 | -5.7273 | -1.6699 | -0.6225 | -0.9914 |
| 31.4 | 0.2390 | -2.9134 | 1.1446 | -1.2851 | -0.4283 | -0.9020 |
| 31.5 | -2.3328 | 4.4380 | 6.9176 | 2.0165 | 1.6176 | 0.1689 |
| 31.6 | -2.1147 | 5.0537 | 4.7025 | -0.4851 | 1.6173 | 0.1854 |
| 31.7 | -1.8350 | 0.2293 | 4.9661 | 1.0755 | 0.6322 | 0.4484 |
| 32.1 | -1.7792 | -0.5736 | 1.3271 | 0.1972 | 1.3385 | 0.6589 |
| 32.2 | -2.7766 | 2.7040 | 0.0689 | -2.0208 | 0.4551 | 0.3902 |
| 32.3 | -2.0369 | -1.6110 | -1.4754 | -0.2839 | 0.4066 | 0.1129 |
| 32.4 | -1.6902 | -9.7029 | -1.7350 | -0.3147 | 1.1100 | 0.1335 |
| 32.6 | -0.9042 | -4.4818 | -0.9866 | -1.6367 | 0.1355 | -0.4134 |
| 32.8 | -0.1031 | -9.3432 | -0.5977 | -0.1401 | -0.8784 | -0.1079 |
| 32.9 | 1.7022 | -2.4418 | -0.0656 | -2.1678 | -0.9110 | -0.3942 |
| 33.0 | -1.6393 | 5.0189 | 5.0646 | 0.1092 | 0.2522 | 0.0300 |

As calculations demonstrate it is sufficient to retain, holding the first few terms of the Fourier series to provide a good approximation and the use of a partial sum which includes the first 5-6 terms occurs sufficient in most cases (see Figures A2.1-A2.5).

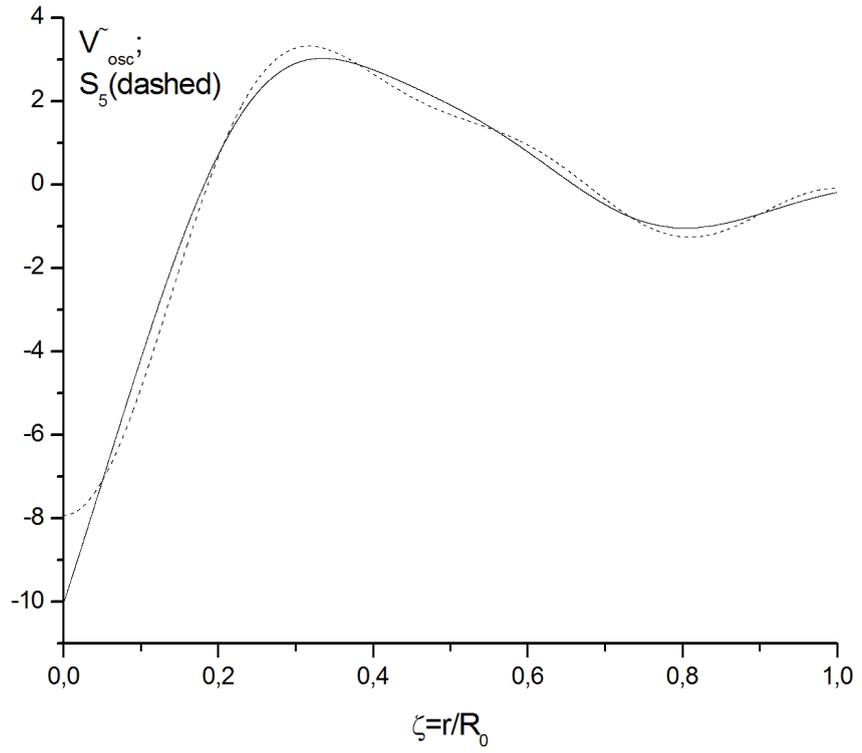

Figure A2.1 Radial dependencies of the averaged oscillation potential and its Fourier series partial sum; $k_F R_0 = 28$

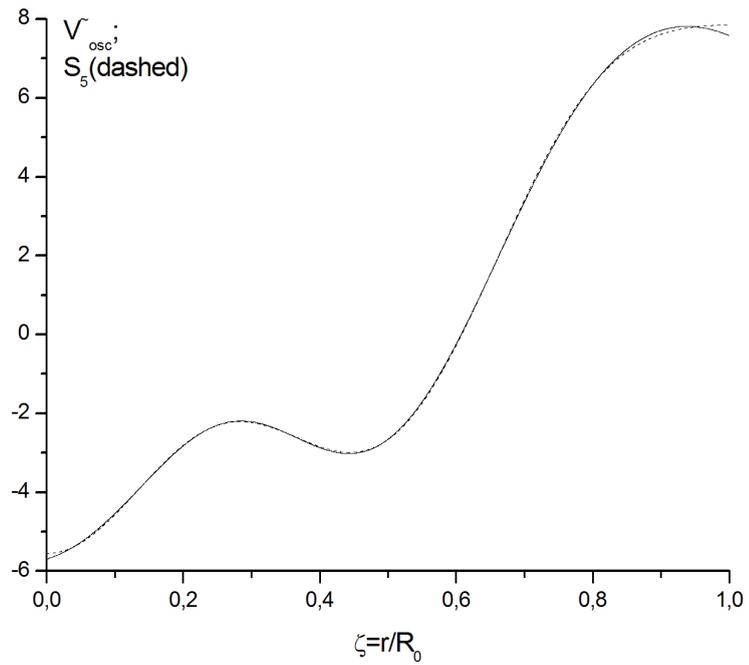

Figure A2.2 Radial dependencies of the averaged oscillation potential and its Fourier series partial sum; $k_F R_0 = 29.4$

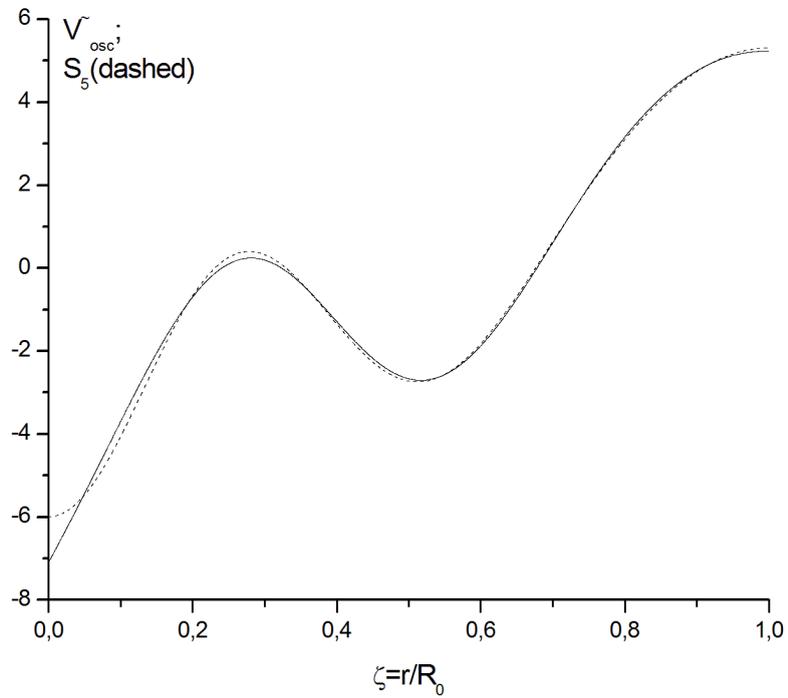

Figure A2.3 Radial dependencies of the averaged oscillation potential and its Fourier series partial sum; $k_F R_0 = 30.8$

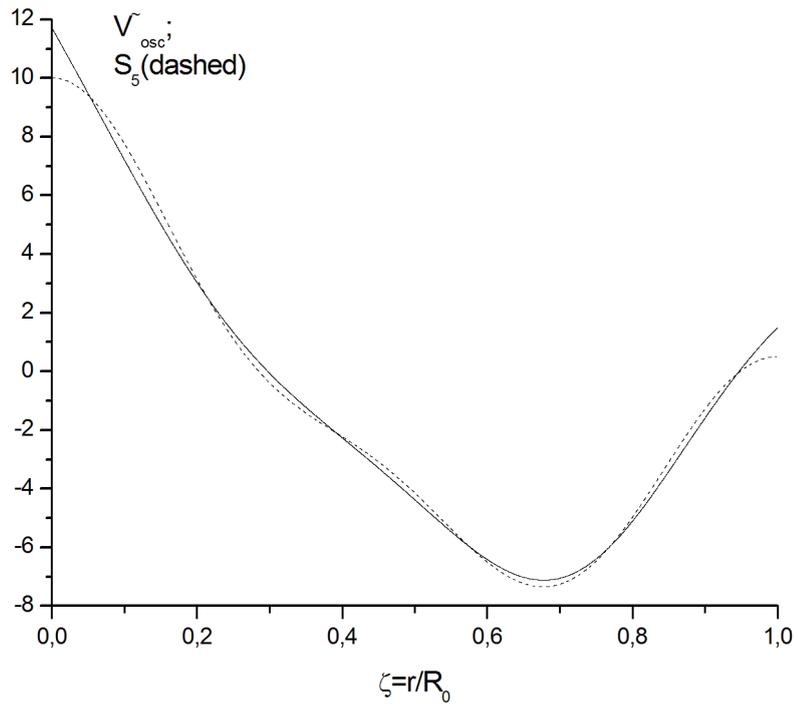

Figure A2.4 Radial dependencies of the averaged oscillation potential and its Fourier series partial sum; $k_F R_0 = 31.6$

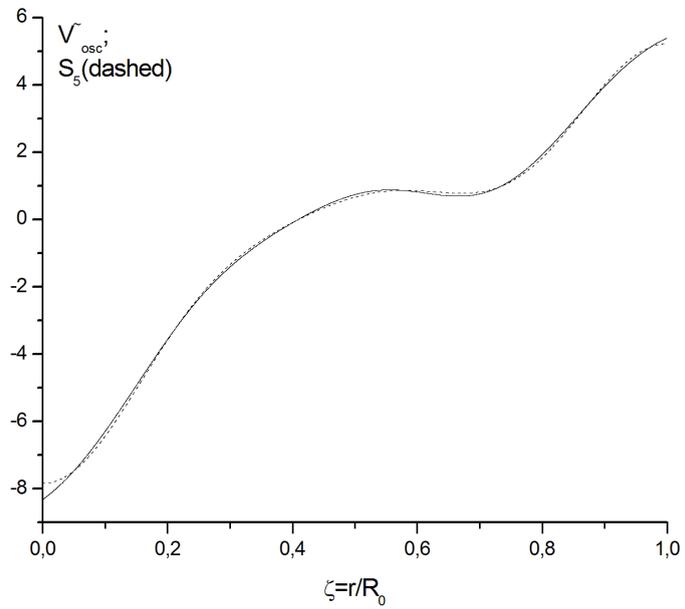

Figure A2.5 $k_F R_0 = 32.6$ Radial dependencies of the averaged oscillation potential and its Fourier series partial sum;

Table A 2.2 shows the values of the coefficients for the barrier potential. The choice of the values of parameter $k_F R_0$ corresponds to the graphics in Fig. 2. A1.7—A1.10.

Table A 2.2 Parameters of the barrier potential for the several values of $k_F R_0$

| $k_F R_0$ | $R_{WS}/R_0$ | $a/R_0$ | $W$ |
|---|---|---|---|
| 28.0 | 9.62E-001 | 1.35E-002 | 439.6 |
| 28.2 | 9.63E-001 | 1.28E-002 | 447.0 |
| 28.3 | 9.63E-001 | 1.30E-002 | 463.9 |
| 28.7 | 9.63E-001 | 1.29E-002 | 471.2 |
| 28.9 | 9.63E-001 | 1.31E-002 | 476.8 |
| 29.2 | 9.64E-001 | 1.22E-002 | 480.3 |
| 29.4 | 9.65E-001 | 1.21E-002 | 489.2 |
| 29.6 | 9.65E-001 | 1.22E-002 | 493.5 |
| 29.7 | 9.65E-001 | 1.24E-002 | 499.9 |
| 29.9 | 9.65E-001 | 1.26E-002 | 515.4 |
| 30.3 | 9.66E-001 | 1.19E-002 | 521.8 |
| 30.6 | 9.66E-001 | 1.16E-002 | 527.2 |
| 30.8 | 9.66E-001 | 1.18E-002 | 532.5 |
| 30.9 | 9.66E-001 | 1.19E-002 | 537.0 |
| 31.1 | 9.66E-001 | 1.20E-002 | 545.7 |
| 31.2 | 9.66E-001 | 1.20E-002 | 551.3 |
| 31.4 | 9.67E-001 | 1.15E-002 | 560.3 |
| 31.5 | 9.67E-001 | 1.17E-002 | 566.2 |
| 31.6 | 9.66E-001 | 1.18E-002 | 571.6 |
| 31.7 | 9.67E-001 | 1.15E-002 | 575.2 |
| 32.1 | 9.67E-001 | 1.16E-002 | 579.9 |
| 32.2 | 9.67E-001 | 1.18E-002 | 585.7 |
| 32.3 | 9.67E-001 | 1.17E-002 | 589.8 |
| 32.4 | 9.68E-001 | 1.10E-002 | 597.5 |
| 32.6 | 9.68E-001 | 1.11E-002 | 602.5 |
| 32.8 | 9.69E-001 | 1.09E-002 | 605.9 |
| 32.9 | 9.68E-001 | 1.11E-002 | 615.9 |
| 33.0 | 9.68E-001 | 1.14E-002 | 622.7 |

# Appendix 3. Numerical algorithm

We consider the sphere of the initial radius $R_0$ consisting of a gas of degenerate electrons with density $n_e(r,t)$ and classical ions with density $n_i$. It is assumed that the electron density satisfies the equilibrium condition at each time moment:

$$-e\frac{\partial}{\partial r}(\varphi + U_{osc} - U_{bar}) + \frac{1}{n_e}\frac{\partial p_e}{\partial r} = 0 \qquad (A3.1)$$

Where $U_{osc}$ and $U_{bar}$ are the oscillating and barrier potentials, respectively:

$$U_{osc}(r) = \sum_{m}^{M} C_m \cos(\pi m \frac{r}{R_0}), \qquad U_{bar}(r) = \frac{W}{1+\exp\left(\frac{r-R_{ws}}{a}\right)}. \qquad (A3.2)$$

where $C$ and $W$ are the characteristic amplitudes of the oscillating and barrier potentials, and $a$ is the size of the barrier potential.

The electron pressure depends on the local electron density and is determined by the following expression:

$$p_e = \frac{(3\pi^2)^{2/3}}{5}\frac{\hbar^2}{m_e}\left(n_e\right)^{5/3}. \qquad (A3.3)$$

The motion of the ion gas is a result of the action on the ions from the electrostatic field created by the instantaneous distribution of ions and electrons. The ion gas is assumed to be ideal, i.e., the effects of viscosity and thermal conductivity are neglected, and is calorically perfect with the equation of state $p=(\gamma-1)\rho\zeta$, where $\gamma=5/3$ is the adiabatic exponent, $p$ is the pressure, and $\zeta$ is the specific internal energy.

It can be shown that the influence of nonequilibrium between the distribution of ions and electrons takes place in a very small neighborhood of the wall. Therefore, in the gas dynamics calculations described below, it is assumed that there is a local equilibrium at all points inside the sphere $n_i = n_e$. Then, the electron pressure determines the ion distribution.

The medium inside the sphere is assumed to be electro-neutral, therefore at each moment the distributions of electrons and ions satisfy the condition of zero total charge:

$$\int_0^{R_0(t)} n_e(r,t) r^2 dr = \int_0^{R_0(t)} n_i(r,t) r^2 dr, \quad n_i = \rho/m_i$$

Where $m_i$ is the mass of ions.

The ion motion is described by the standard system of Euler equations for gas dynamics under the assumption of spherical symmetry. This system is supplemented by the right-hand side corresponding to the change in momentum and energy due to the electrostatic field produced by charged particles in the gas (ions and free electrons):

$$\begin{cases} \dfrac{\partial \rho}{\partial t} + \dfrac{\partial (\rho U)}{\partial r} = -\dfrac{2}{r}\rho U, \\[4pt] \dfrac{\partial (\rho U)}{\partial t} + \dfrac{\partial (\rho U^2 + p)}{\partial r} = -\dfrac{2}{r}\rho U^2 - \dfrac{e}{m_i}\rho\dfrac{\partial \varphi}{\partial r}, \\[4pt] \dfrac{\partial (\rho E)}{\partial t} + \dfrac{\partial (\rho U H)}{\partial t} = -\dfrac{2}{r}\rho U H - \dfrac{e}{m_i}\rho U\dfrac{\partial \varphi}{\partial r}, \end{cases}$$

$$(A3.4)$$

where e is the electron charge, $E = 0.5U^2 + \zeta$ is the specific total ion energy, $H = \zeta + p/\rho$ is the specific total enthalpy, $\zeta = p/\rho/(\gamma-1)$ is the specific internal energy of the ions, $\gamma = 5/3$, and $0 \leq r \leq R(t)$. The potential is determined by the equation (A3.1), where we use $n_e$ instead: $n_i$.
We rewrite the system of equations (A3.1), (A3.4) in dimensionless form. To do this, we introduce the scales of the characteristic physical quantities in the following form:

$\rho_* = 10^3$ kg/m$^3$ - density

$L_* = 10^{-6}$ m - length

$U_* = 10^3$ m/sec - velocity

$t_* = L_*/U_* = 10^{-9}$ sec - time

$p_* = \rho_* U_*^2 = 10^9$ N/m$^2$ - pressure

$\varphi_* = m_i U_*^2/e$ V $= 1.875 \times 10^{-2}$ V - potential of electric field

$\Theta_* = m_i U_*/e\, t_* = 1.875 \times 10^4$ V/m - electric field

The values of the physical parameters of the model are:

$e = 1.6 \times 10^{-19}$ Cl - electron charge

$m_i = 3 \times 10^{-27}$ kg – ion mass

The system of equations (A3.4) can be written in a conservative form. In dimensionless variables, it has the following form:

$$\begin{cases} \dfrac{\partial r^2 \rho}{\partial t} + \dfrac{\partial (r^2 \rho U)}{\partial r} = 0, \\ \dfrac{\partial (r^2 \rho U)}{\partial t} + \dfrac{\partial r^2(\rho U^2 + p)}{\partial r} = 2rp - \rho r^2 \dfrac{\partial \varphi}{\partial r}, \\ \dfrac{\partial r^2(\rho E)}{\partial t} + \dfrac{\partial r^2(\rho U H)}{\partial t} r = -\rho U r^2 \dfrac{\partial \varphi}{\partial r}, \end{cases}$$

(A3.5)

where the gradient of the potential

$$\dfrac{\partial \varphi}{\partial r} = K_1 \rho^{-1/3} \dfrac{\partial \rho}{\partial r} + K_3 \left[\dfrac{R_0(0)}{R_0(t)}\right]^2 \dfrac{\partial U_{bar}}{\partial r} - K_4 \dfrac{R_0(0)}{R_0(t)} \dfrac{\partial f(r, R_0(t))}{\partial r}.$$

The dimensionless constants are as follows:

$$K_1 = \dfrac{(3\pi^2)^{2/3}}{3} \dfrac{\hbar^2}{em_e \varphi_*} (n_*)^{2/3}, \quad n_* = \rho_*/m_i$$

$$K_3 = \dfrac{\hbar^2}{2em_e \varphi_*} \left(3\pi^2 \dfrac{\rho_0}{m_i}\right)^{2/3} \left[\dfrac{R_0(0)}{R_0(t)}\right]^2,$$

$$K_4 = C \dfrac{10^{-10} m}{\varphi_*} \left(\dfrac{\rho_0}{m_i}\right)^{1/3}$$

If we introduce the total pressure

$$\pi = p + \dfrac{3}{5} K_1 (\rho)^{5/3},$$

the system of equations (A3.5) can be rewritten as follows:

$$\begin{cases} \dfrac{\partial r^2 \rho}{\partial t} + \dfrac{\partial (r^2 \rho U)}{\partial r} = 0, \\ \dfrac{\partial (r^2 \rho U)}{\partial t} + \dfrac{\partial r^2 (\rho U^2 + \pi)}{\partial r} - 2r\pi = -\rho r^2 \left\{ K_3 \left[ \dfrac{R_0(0)}{R_0(t)} \right]^2 \dfrac{\partial U_{bar}}{\partial r} - K_4 \dfrac{R_0(0)}{R_0(t)} \dfrac{\partial f(r, R_0(t))}{\partial r} \right\}, \\ \dfrac{\partial r^2 (\rho E)}{\partial t} + \dfrac{\partial r^2 \rho U(E + \pi/\rho)}{\partial r} = \dfrac{3}{5} K_1 (\rho)^{5/3} \dfrac{\partial r^2 u}{\partial r} - \\ \qquad - \rho r^2 U \left\{ K_3 \left[ \dfrac{R_0(0)}{R_0(t)} \right]^2 \dfrac{\partial U_{bar}}{\partial r} - K_4 \dfrac{R_0(0)}{R_0(t)} \dfrac{\partial f(r, R_0(t))}{\partial r} \right\}, \end{cases}$$

(A3.6)

Except for the right-hand side, the system of equations (A3.6) coincides exactly with the classical system of equations of gas dynamics with the equation of state (EOS) in the following form:

$$\pi = (\gamma - 1)\rho e + \dfrac{3}{5} K_1 (\rho)^{5/3}$$

(A3.7)

The numerical integration of (A3.6) can be performed by the Godunov method,

$$\mathbf{q}_i^{n+1} = \dfrac{3}{\left(r_{i+1/2}^{n+1}\right)^3 - \left(r_{i-1/2}^{n+1}\right)^3} \left\{ \dfrac{\left(r_{i+1/2}^n\right)^3 - \left(r_{i-1/2}^n\right)^3}{3} \mathbf{q}_i^n - \Delta t \left[ \left(r_{i+1/2}^{n+1/2}\right)^2 \mathbf{F}_{i+1/2} - \left(r_{i-1/2}^{n+1/2}\right)^2 \mathbf{F}_{i-1/2} + \mathbf{S}_{\pi,i} + \mathbf{S}_{U,i} \right] \right\}$$

here $\mathbf{F}$ is denotes the fluxes in divergence terms of the system (3.6), and

$$\mathbf{S}_{\pi,i} = \begin{bmatrix} 0 \\ \left( \left(r_{i+1/2}^{n+1/2}\right)^2 - \left(r_{i-1/2}^{n+1/2}\right)^2 \right) \pi_i \\ 0 \end{bmatrix},$$

$$\mathbf{S}_{U,i} = \begin{bmatrix} 0 \\ \left( \left(r_{i+1/2}^{n+1/2}\right)^3 - \left(r_{i-1/2}^{n+1/2}\right)^3 \right) \rho_i S_{2,i} \\ \left( \left(r_{i+1/2}^{n+1/2}\right)^2 U_{i+1/2} - \left(r_{i-1/2}^{n+1/2}\right)^2 U_{i-1/2} \right) 3 K_1 (\rho_i)^{5/3} / 5 + \left( \left(r_{i+1/2}^{n+1/2}\right)^3 - \left(r_{i-1/2}^{n+1/2}\right)^3 \right) \rho_i U_i S_{3,i} \end{bmatrix}$$

(A3.8)

Here, the right-hand side of the equations is related with the barrier and oscillating potential, and the values at the edges of the cells (half-integral indices) are determined from the solution of the Riemann problem. To solve the Riemann problem, we use the local approximation of the EOS (A3.7)

$$\pi = (\gamma - 1)\rho e + c_0^2 (\rho - \rho_0),$$

(A3.9)

where the parameters of the EOS are approximated from the value of the local density by the following relations:

$$\rho_0 = \dfrac{2}{5} \rho, \qquad c_0^2 = K_1 \rho^{2/3}$$

(A3.10)

The Courant number is determined by the speed of sound of the system (A3.6) to provide stability of the numerical scheme.

$$c^2 = \frac{\gamma(\pi+\pi_0)\rho e + \frac{3}{5}K_1(\rho)^{5/3}}{\rho}, \quad \pi_0 = \frac{2}{5\gamma}K_1(\rho)^{5/3} \qquad (A3.11)$$